\documentclass[10pt,preprint]{aastex}

\usepackage{amsmath}
\usepackage{amssymb}
\usepackage{amsthm}
\usepackage{graphicx}
\usepackage{natbib,color,lscape}
\usepackage{multirow}
\usepackage{epsfig}
\usepackage[nofiglist,notablist,noheads,nomarkers]{endfloat}

\newcommand{\sm}{2M0535$-$05}
\newcommand{\pr}{Par~1802}
\newcommand{\ic}{$I_C$}
\newcommand{\op}{$VI_C$}
\newcommand{\prv}{Paper~I}
\newcommand{\pii}{Paper~II}
\newcommand{\nir}{{\it JHK$_S$}}
\newcommand{\phoebe}{{\tt PHOEBE}}
\newcommand{\teff}{$T_{\rm eff}$}
\newcommand{\msun}{M$_\odot$}

\shorttitle{Parenago 1802: Dissimilar PMS Twins}
\shortauthors{G\'omez Maqueo Chew et al.}

\begin{document}

\title{Luminosity Discrepancy in the Equal-Mass, Pre--Main Sequence Eclipsing Binary Par~1802: Non-Coevality or Tidal Heating?}

\author{
Yilen G\'omez Maqueo Chew\altaffilmark{1,2},
Keivan G.\ Stassun\altaffilmark{1,3,4},
Andrej Pr\v{s}a\altaffilmark{5,6},
Eric Stempels\altaffilmark{7},
Leslie Hebb\altaffilmark{1},
Rory Barnes\altaffilmark{8},
Ren\'{e} Heller\altaffilmark{9},
Robert D.\ Mathieu\altaffilmark{10}
}

\altaffiltext{1}{Department of Physics and Astronomy, 
Vanderbilt University, Nashville, TN 37235, USA; 
yilen.gomez@vanderbilt.edu}
\altaffiltext{2}{Astrophysics Research Centre, Queen's University Belfast, 
University Rd.\ Belfast, County Antrim BT7 1NN, UK}
\altaffiltext{3}{Department of Physics, Fisk University, Nashville, TN 37208, USA}
\altaffiltext{4}{Department of Physics, Massachusetts Institute of Technology, Cambridge, MA 02139, USA}
\altaffiltext{5}{Department of Astronomy and Astrophysics, Villanova University, Villanova, PA 19085, USA}
\altaffiltext{6}{Department of Physics, University of Ljubljana, 1000 Ljubljana, Slovenia}
\altaffiltext{7}{Department of Astronomy and Space Physics, Uppsala University, SE-752 67 Uppsala, Sweden}
\altaffiltext{8}{Department of Astronomy, University of Washington, Seattle, WA 98195, USA}
\altaffiltext{9}{Leibniz-Institut f\"ur Astrophysik Potsdam (AIP), An der Sternwarte 16, 14482 Potsdam, Germany}
\altaffiltext{10}{Department of Astronomy, University of Wisconsin---Madison, Madison, WI 53706, USA}

\begin{abstract}
\object{Parenago 1802}, a member of the $\sim$1 Myr Orion Nebula Cluster,
is a double-lined, detached eclipsing binary in a 4.674~d orbit, 
with equal-mass components ($M_2/M_1$=0.985$\pm$0.029).
Here we present extensive \op\nir\ light 
curves spanning $\sim$15~yr, as well as a Keck/HIRES optical spectrum. 
The light curves evince a third light source
that is variable with a period of 0.73~d, and is also manifested
in the high-resolution spectrum, strongly indicating the presence of a third star in 
the system, probably a rapidly rotating classical T~Tauri star.
We incorporate this third light into our radial velocity and light curve modeling 
of the eclipsing pair, measuring 
accurate masses ($M_1$=0.391$\pm$0.032, $M_2$=0.385$\pm$0.032 M$_\sun$),
radii ($R_1$=1.73$\pm$0.02, $R_2$=1.62$\pm$0.02 R$_\sun$),
and temperature ratio
($T_{\rm eff,1}/T_{\rm eff,2}$=1.0924$\pm$0.0017).
Thus the radii of the eclipsing stars differ by 6.9$\pm$0.8\%, 
the temperatures differ by 9.2$\pm$0.2\%,
and consequently the luminosities differ by 62$\pm$3\%,
despite having masses equal to within 3\%.
This could be indicative of an age difference of $\sim3\times10^5$ yr between
the two eclipsing stars, perhaps a vestige of the binary formation history. 
We find that the eclipsing pair is  
in an orbit that has not yet fully circularized, $e$=0.0166$\pm$0.003.  
In addition, we measure the rotation rate of the
eclipsing stars to be 4.629$\pm$0.006~d; they rotate slightly
faster than their 4.674~d orbit. 
The non-zero eccentricity and super-synchronous rotation suggest that
the eclipsing pair should be tidally interacting, so we calculate
the tidal history of the system according to different 
tidal evolution theories. We find that tidal heating effects 
can explain the observed 
luminosity difference of the eclipsing pair, providing an alternative to the previously suggested age difference. 
\end{abstract}


\section{Introduction}

The initial mass and chemical composition of newly formed stars are
key factors in determining their evolutionary path. Multiple systems are
commonly considered to be formed simultaneously from the same protostellar
core, such that their components are assumed to be coeval and to have the
same metallicity.  Equal-mass components of binary systems---i.e.,
{\it twins} \citep[e.g.,][]{simon09}---are therefore expected to evolve 
following essentially the same evolutionary track.

Eclipsing binary (EB) systems are useful observational tools that 
render direct measurements of their components' physical parameters, 
independent of theoretical models and distance 
determination, against which theoretical evolutionary models can be tested.  
There are a few tens of pre--main-sequence (PMS)
systems for which the dynamical stellar masses 
are measured \citep[][and references therein]{mat07}; 
EBs however are the only ones 
that allow for the  direct measurement of the radii of the components. 
EBs are rare, because their orbits have to be 
oriented such that we see the components eclipse.  
For PMS, low-mass EBs, where both components have masses 
below 1.5 M$_\sun$, there are only seven such systems reported in the 
literature: 
ASAS~J052821+0338.5 \citep{asas08}; 
RX~J0529.4+0041 \citep{cov00,cov04}; 
V1174~Ori \citep{sta04}; 
MML~53 \citep{hebb10};
Parenago~1802 \citep[][and target of this study]{rvsm2654,natsm2654}; 
JW~380 \citep{jw380}, 
and \sm\ \citep{nt06,pii,sm4147}. 
For the latter, the components are below the hydrogen-burning limit, 
i.e., they are brown dwarfs. For this particular system, 
the effective temperatures of the two bodies are observed to be reversed 
with the more massive brown dwarf appearing to be cooler. 
Similar to the approach we apply here for \pr, 
\citet{Heller2010} have explored the effects of tidal heating in that system. 

	The discovery of \pr\ was previously presented, along with its radial velocity 
	study that found the system to be an EB with a period of $\sim$4.67~d 
	where both components have near equal masses,  
	$M_1$ = 0.40$\pm$0.03 \msun\ and $M_2$ = 0.39$\pm$0.03 \msun\ 
        \citep[][hereafter \prv]{rvsm2654}. 
	\pr, as a member of the Orion Nebula Cluster \citep[ONC;][]{hill97}, 
	is considered to have an age of $\sim$1 Myr (\prv).
	A follow-up analysis which included the radial velocity curves and 
	the \ic-band light curve found the components' masses to be equal to within 
	$\sim$2\%, but their radii and effective temperatures to differ by
	$\sim$5--10\% \citep[][hereafter \pii]{natsm2654}.  
	They suggest that these disparate radii and temperatures are
	the result of a difference in age of a few hundred thousand years.

	In this paper, we present new \op\nir\ light curves for \pr\ 
as well as
	a newly acquired high-resolution optical spectrum (\S\ref{data}). 
	The multi-band nature of our analyses (\S\ref{analysis}) allows us to probe the radiative properties of the system. The analysis includes
	an in-depth periodicity analysis of the light curves, which enables us to refine the orbital 
	period for the binary and identify the rotation periods of its components (\S\ref{period}).  
	We are also able to measure the presence of a third light source
        in the system (\S\ref{thirdstar}), 
        through identification of a very short-period modulation in
        the light curves that definitively cannot be due to rotation of 
        either of the eclipsing stars,
through the analysis of an additional continuum contribution in the spectra, 
	through analysis of third-light dilution in the light curves, 
        and through analysis of the system's broadband spectral energy distribution. 
        We combine these analyses into a comprehensive, global model of the
        EB's fundamental orbital and physical properties (\S\ref{eb}),
        along with formal and heuristic uncertainties (obtained from a direct $\chi^2$ mapping of the parameter space) 
        in these parameters. 
        \pr\ is found to be a low-mass PMS EB with a nominal age of 
        $\sim$1~Myr, comprising two equal-mass eclipsing stars of 
        0.39~M$_\sun$ and a third similarly low-mass star, 
        probably in a wide orbit, that is rapidly
        rotating and likely accreting (i.e., a Classical T~Tauri star).
The radii of the eclipsing pair differ by 6.9$\pm$0.8\%, 
their effective temperatures differ by 9.2$\pm$0.2\%,
and consequently, their luminosities differ by 62$\pm$3\%,
despite their masses being equal to within 3\%.

        In \S\ref{diss}, we discuss possible explanations for the
        large difference in luminosity of the eclipsing pair, including magnetic activity,  
        non-coevality arising from mass-equalizing effects in the 
        binary's formation, and 
	tidal heating arising from the binary's past orbital evolution.
The last two explanations appear plausible, with the latter predicting a
possible misalignment of the stellar spin axes, which could be observable.
We summarize our conclusions in \S\ref{summary}.

	\section{Data\label{data}}

	\subsection{Photometric Observations \label{lcs}}

	We present the light curves of \pr\ in $V$ (with a total of 2286 data points), \ic\ (3488), $J$ (564), $H$ (176) and $K_S$ (365).  The detailed observing campaign is described in Table \ref{table:obs}, and the individual measurements in each observed passband are given in Tables \ref{table:lcv}--\ref{table:lck}.  The \ic\ data cover the largest time span, from December 1994 to January 2009; it includes the previously published light curve (\pii) and 1279 new data points obtained between March 2007 and January 2009. The $V$ light curve includes data obtained between January 2001 and January 2009 with the 0.9-m telescope at KPNO and with the SMARTS 0.9-m, 1.0-m and 1.3-m telescopes at CTIO.  
	Using the ANDICAM instrument which allows for simultaneous optical and near-infrared imaging, \pr\ was observed photometrically with the SMARTS 1.3-m telescope at CTIO between February 2005 and February 2008, constituting the entirety of the \nir\ light curves.
	We also observed \pr\ in the $B$-band; however, the resulting light curve was not 
	well-sampled and it is very noisy due to the photometric variability of the third star in the system (see below).
	Thus, we do not include the $B$-band in the rest of our analyses, except 
	as a consistency check of our final solution.

Because the light curve data were obtained mostly in queue mode on a variety
of instruments over a long period of time, individual exposure times varied
depending on the instrument and observing conditions. Typically, however,
the $VI_C$ light curve data were obtained with 
typical exposure times of 60~s.
	The observations in the near-infrared were made in sets of five
	dither positions for the $J$-band, and seven dither positions for
	the $HK_S$-bands, with total integration times of 150~s and 175~s,
	respectively.

The optical (\op) light curves were determined via ensemble differential PSF photometry 
(see \citet{honey92,sta99,sta02}; and references therein)
using the full ensemble of other stars in the field view, 
which is typically several hundred stars, except in the ANDICAM data for which it is typically a few tens of stars. Thus the differential light curve solutions do not rely upon individual comparison stars, and the solutions are much more robust against CCD-wide systematics. 
Data from different seasons and/or instruments are placed onto a common photometric scale determined from this large ensemble of comparison stars. 
In addition, to allow for small systematic offsets from season to season or across instruments, we subtracted from each season's light curve the median out-of-eclipse value, which is well determined because of the large number of data points in each light curve. 

	Differential photometry on the $JHK_S$ light curves 
        followed the procedures in \citet{sm4147}, using an aperture of 6 pixels, which 
corresponds to 1.5 times the typical FWHM of the images.  
	In this case, the comparison star used for the $JHK_S$ light curves was Par~1810,
	chosen because it is present in all of the reduced images of \pr\ and 
        because it exhibits no
	variability in the \ic\ and $V$ bands; furthermore, it is not listed as 
	variable in the near-infrared variability study of the ONC by \citet{carp}.
The typical uncertainty in the $VI_C$ light curves is $\sigma_I = 0.01$ mag
        and $\sigma_V = 0.025$ mag, the latter dominated by 
        poor sky subtraction due to scattered light in the nebular background.
	The uncertainty in the produced \nir\ light curves is dominated by the
	systematic uncertainties in the sky background subtraction.
	The $JH$ bands have a similar scatter, $\sigma_J =
	\sigma_H = 0.01$ mag; however, the interference pattern of the sky emission
	lines in the $K_S$ light curve is more significant making the scatter in
	this band larger, $\sigma_{K_S} = 0.02$ mag.
	These uncertainties were estimated by calculating the standard 
        deviation of the light curves, with the data during eclipses 
        excluded and the periodic low-amplitude variability 
        (see \S\ref{period}) subtracted.

\begin{figure}[tbp]
\includegraphics{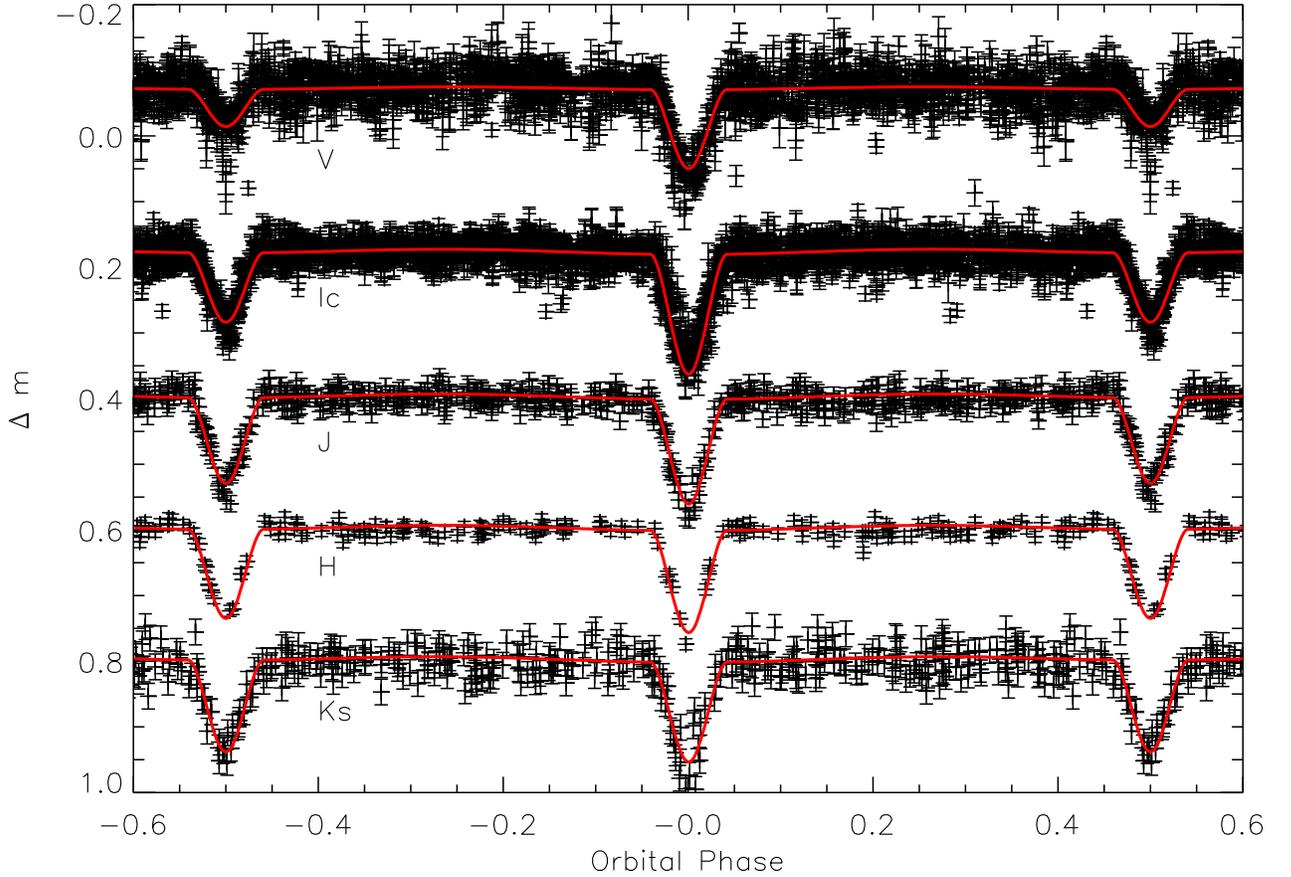}
\caption[Observed and Modeled \op\nir\ Light Curves of \pr.]{\label{fig:lcs}
Observed and Modeled \op\nir\ Light Curves of \pr. 
We show the observed photometric light curves with their corresponding uncertainties
as described in \S\ref{lcs}.  The data have been folded over the binary's orbital period 
and shifted in magnitude for easier visualization.
The solid line
represents the best RV+LC solution for \pr\ (see \S\ref{eb} for 
a detailed description of the modeling procedure,
and see Table \ref{table:params} for the physical parameters of the EB components and their orbit). 
}
\end{figure}

Fig.~\ref{fig:lcs} shows the \op\nir\ light curves, 
	including those published in \pii. The data
	have been folded on the orbital period and each band has been 
        offset for clarity.  
	Each point is an individual observation and the solid line represents the model of our final light-curve solution
	(\S\ref{eb}).

	\subsection {Spectroscopic Observations \label{hires}}

We observed \pr\ on the night of UT 2007 Oct 23 with the
High Resolution Echelle Spectrometer (HIRES) on
Keck-I\footnote{Time allocation through NOAO via the NSF's
Telescope System Instrumentation Program (TSIP).}.
The exposure time was 900~s.
We observed in the spectrograph's ``red" (HIRESr) configuration
with an echelle angle of $-0.403$ and a cross-disperser angle of 1.703.
We used the OG530 order-blocking filter and 1\farcs15$\times$7\farcs0
slit, and binned the chip during readout by 2 pixels in the dispersion direction.
The resulting resolving power is $R\approx 34\, 000$ per 3.7-pixel
($\sim 8.8$ km s$^{-1}$) FWHM resolution element.
For the analyses discussed below, we used the 21 spectral orders from the
``blue" and ``green" CCD chips, covering the wavelength range
$\lambda\lambda$5782--8757.
ThAr arc lamp calibration exposures were obtained before and after
the \pr\ exposure, and sequences of bias and flat-field exposures
were obtained at the end of the night. The data were processed using standard
IRAF\footnote{IRAF is distributed by the National Optical Astronomy
Observatory, which is operated by the Association of Universities for
Research in Astronomy (AURA) under cooperative agreement with the National
Science Foundation.}
tasks and the MAKEE reduction package written for HIRES
by T.~Barlow, which includes optimal extraction of the
orders as well as subtraction of the adjacent sky background.
The signal-to-noise of the final spectrum is $\approx 70$
per resolution element.

In addition, we observed the late-type spectral
standards \citep[see][]{kirkpatrick91}, Gl~205 (M1) and Gl~251 (M3),
at high signal-to-noise. These spectral types were chosen 
to match the inferred spectral types of the eclipsing components of \pr, 
based on the tomographic reconstruction analysis presented in \pii.
They were observed immediately before the \pr\  
exposure and used exactly the same instrumental configuration. We use these
spectral standards below in our spectral decomposition analysis of the \pr\  
spectrum.

	\section{Analysis of Periodicity and Third Light\label{analysis}}

	\subsection{Periodicities in the \pr\ Light Curves \label{period}}

	We measure the timings of the eclipses in the \ic\ light curve, which covers the longest time span,
	 and are able to refine the ephemeris for \pr\ by 
	 performing a least-square fit to the observed eclipse times.
	The individual eclipse times are measured by a least-squares fit of a Gaussian 
	to those eclipses for which there are at least five data points and that include the minimum of each eclipse. 
	 Table \ref{table:timings} summarizes the measurements of the timings of the eclipses and their uncertainties.  
	We find a best-fit orbital period of 
$P_{\rm orb}$=4.673903$\pm$0.000060~d, 
 and epoch of primary minimum HJD$_0$=2454849.9008$\pm$0.0005~d,
which we adopt throughout our analysis as the system's ephemeris.

The eclipse times in Table~\ref{table:timings} show an r.m.s.\ 
scatter of 28~min, which is much larger than the formal uncertainty 
of a few seconds on the individual timing measurements. 
We have checked for systematic trends in these timing variations on
timescales of 1--20~yr, such as might be produced by reflex motion of
the eclipsing pair induced by the third body in the system (see
\S\ref{thirdstar}). However, we do not find evidence for systematic deviations 
of the eclipse times from the above simple linear ephemeris. 
Instead, we regard the scatter as more likely arising from spots on the stars, as manifested in the periodic light curve modulation from which we measure the stellar rotation period (see below). Surface spots can induce asymmetry in the eclipses and thus effectively shift the eclipse minima by a small fraction of the star crossing time \citep[e.g.,][]{torr02,sta04}. Indeed, the 28~min scatter in eclipse times corresponds to 0.0041 orbital phase, which is a small fraction ($\sim$4\%) of the eclipse duration. As an example, adding a single cool spot (50\% cooler than the photosphere) covering 1\% of the primary star's surface, can induce a shift of 20~min in the time of primary eclipse predicted by our non-spotted light curve model (see Sec.~\ref{eb}). Spots on the secondary eclipsing star, as well as the short-period light curve variations that we see from the rapidly rotating third star (see below), likely introduce additional shifts of comparable magnitude in the observed eclipse timings.

\begin{figure}[tbp]
\includegraphics[trim=0.1in 3.2in 0.05in 0.05in,clip=true]{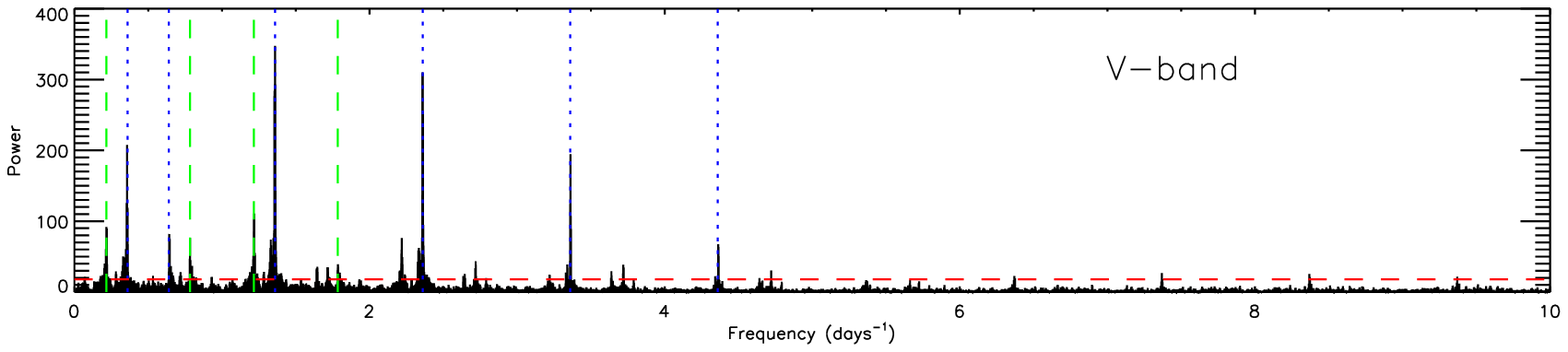}
\includegraphics[trim=0.1in 0.05in 0.05in 1.8in,clip=true]{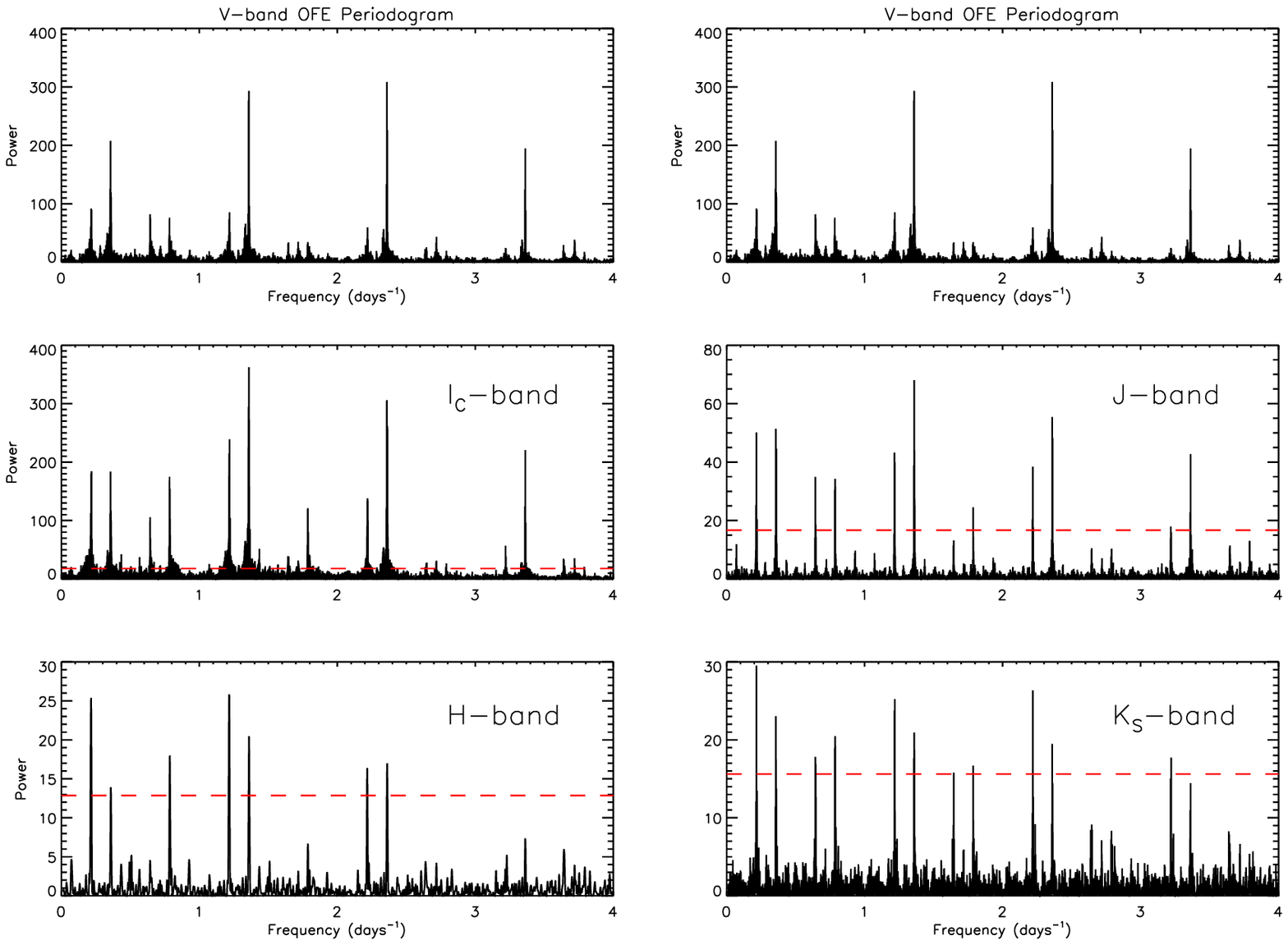}
\caption[OFE \op\nir\ Lomb-Scargle Periodograms.]{\label{fig:periodogram}
OFE \op\nir\ Lomb-Scargle Periodograms.  
The out-of-eclipse (OFE) light curves were searched 
for periodicities, as described in \S\ref{period}, 
identifying two independent periodic 
signals with frequencies of $\sim$0.216  and $\sim$1.36 d$^{-1}$,
corresponding to periods of $P_{1} = 4.629 \pm 0.006$ 
and $P_{2} = 0.7355 \pm 0.0002$ d, respectively.  Table \ref{table:periods} lists the identified periods in each observed
passband with their corresponding uncertainties.  
The vertical, dashed lines on the top panel mark the frequency corresponding to $P_{1}$ and its
aliases and beats; while the vertical, dotted lines correspond to the frequency of $P_{2}$ and its aliases and beats.
The significance of the peaks is given by the horizontal, dashed line which 
denotes the 0.1\% False-Alarm Probability (FAP); since most of the significant peaks are 
found between 0 and 4 d$^{-1}$, only the $V$-band periodogram is shown in its entirety. 
The out-of-eclipse \op\nir\ light curves folded over the two 
identified periods are presented in Fig. \ref{fig:phlcs}. }
\end{figure}

The \op\nir\ light curves corresponding to the
out-of-eclipse (OFE) phases, i.e., all phases excluding those during the eclipses,
are searched for periods between 0.1 and 20 d using
the Lomb-Scargle periodogram technique \citep{scargle}, 
which is well suited to our unevenly sampled data.
The resulting periodograms (Fig.\ \ref{fig:periodogram}) show the power
spectra in frequency units of d$^{-1}$ and present multiple strong peaks.
These peaks represent a combination of one or more true independent signals 
and their aliases.

The amplitudes of the periodograms are normalized 
by the total variance of the data \citep{horne},
yielding the appropriate statistical behavior which allows
for the calculation of the false-alarm probability (FAP).
To calculate the FAPs for each of the OFE light curves, 
a Monte Carlo bootstrapping method \citep[e.g.,][]{sta99} is
applied; it does 1000 random combinations of the differential
magnitudes, keeping the Julian Dates fixed in order to preserve the
statistical characteristics of the data.
The resulting 0.1\% FAP level is indicated in the periodograms by the horizontal dashed line in
Fig.\ \ref{fig:periodogram}. All periodogram peaks higher than the 0.1\% FAP are considered
to be due to real periodicity in our data; this includes the aliases and beats of any periodic signals.

\begin{figure}[tbp]
\includegraphics[angle=90,scale=0.75,trim=0 0.5 0 4.25in,clip=true]{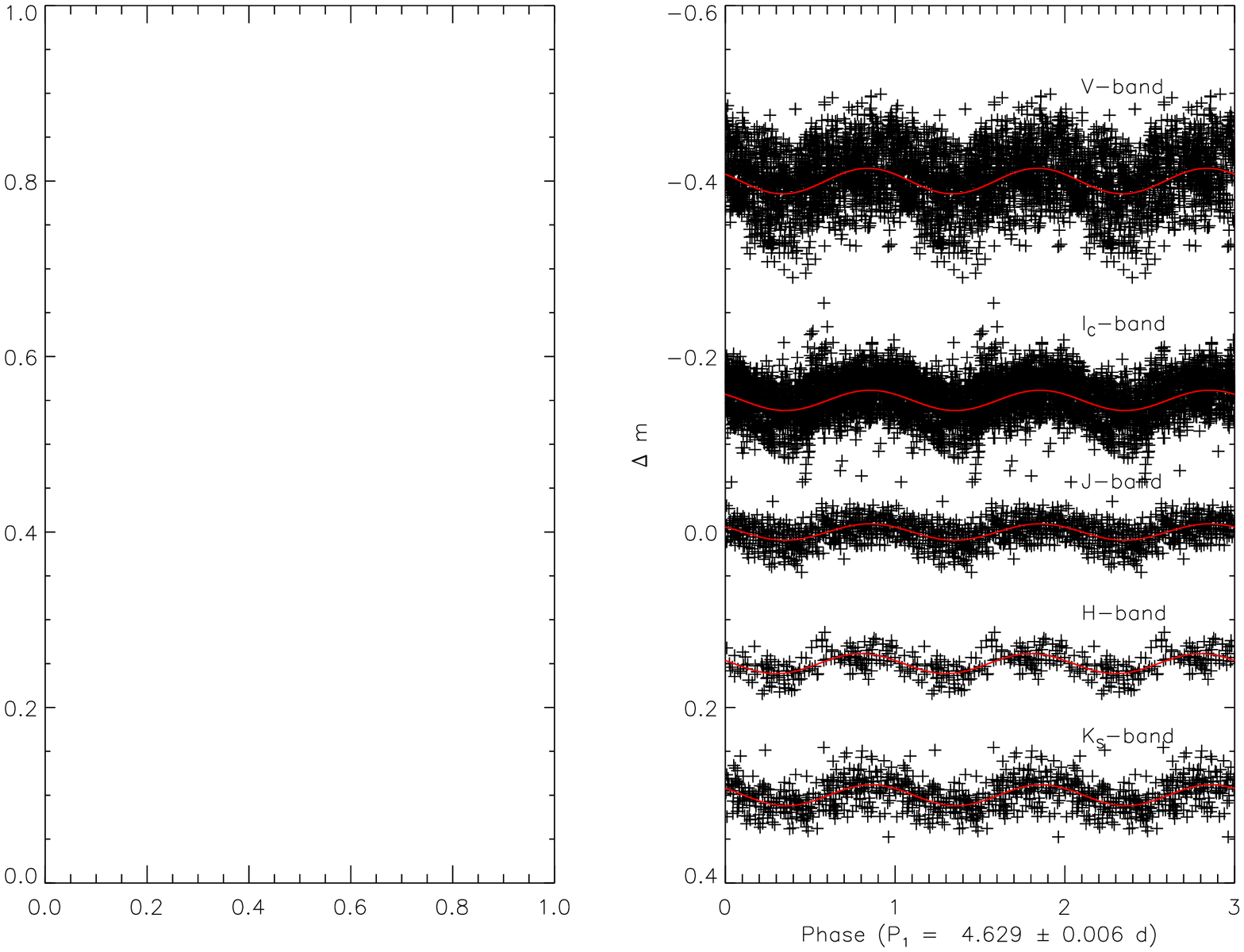}
\includegraphics[angle=90,scale=0.75,trim=0 0.3 0 4.55in,clip=true]{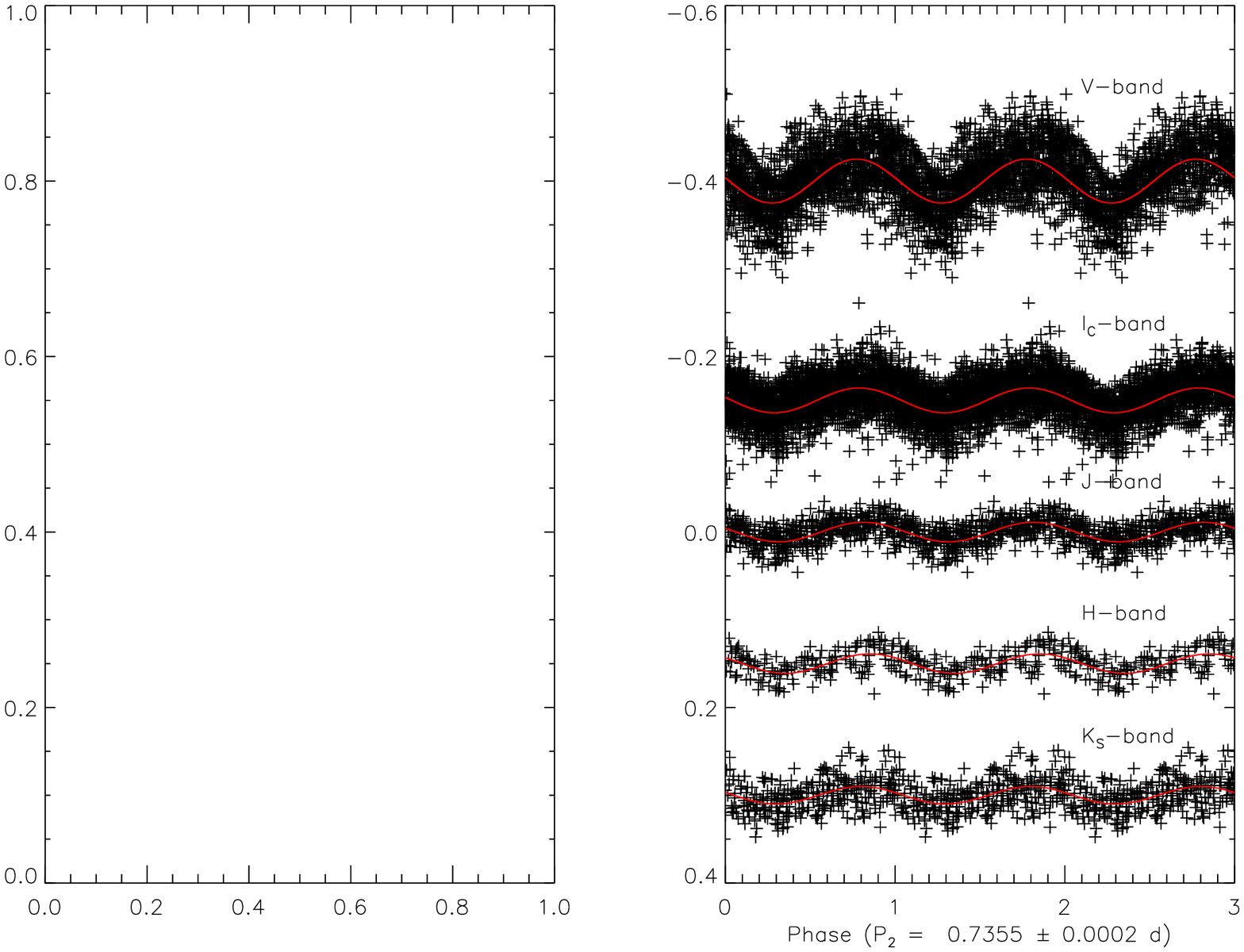}
\caption[Low-Amplitude, Photometric Variability.]{\label{fig:phlcs}
Low-Amplitude, Photometric Variability.
The sinusoidal shape shown by the OFE light curves, folded over either of the two independent periods found
 in all observed passbands from the periodicity analysis (see Fig. \ref{fig:periodogram} and Table \ref{table:periods}), 
is characteristic of spot-induced, rotational modulation. 
The left-hand panel shows the \op\nir\ light curves folded over $P_1$ and displaced 
from zero for easier visualization.  Superimposed is a sinusoid of period $P_1$ fitted to the data.  
In a similar way, the right-hand panel shows 
the same photometric OFE data folded over the shorter period, $P_2$, and its corresponding sinusoidal fit.  
The actual data points are repeated over each of the three phases shown. 
$P_1$ is attributed to the rotation period of the eclipsing components, and is consistent
with their measured $v\sin{i}$ and radii; whereas $P_2$ is attributed
to the stellar source of third light (see \S\ref{thb} for discussion on the third body).
The amplitudes of this spot-induced variation at different passbands
are obtained from the simultaneous fit of two sinusoids, and are given in Table~\ref{table:amps}.
}
\end{figure}

To distinguish the periodogram peaks of the independent periods from their aliases, a sinusoid is
fitted to each light curve and subtracted from the data in order to remove the
periodicity corresponding to the strongest peaks in the periodograms.
This filtering procedure allows us to identify in the OFE periodograms 
of all observed passbands two independent periods, 
$P_1$ = 4.629 $\pm$ 0.006 days and $P_2$  = 0.7355 $\pm$ 0.0002 days.
These two periods are given by the mean of the individual period measurements in each band 
and their uncertainties are given by the standard deviation of the mean 
(see Table \ref{table:periods}).
When the OFE light curves  
are phased to either $P_1$ or $P_2$, they are characterized by having a sinusoidal low-amplitude variability 
which is indicative of stellar rotational modulation \citep[e.g.,][]{sta99}. 
Fig.~\ref{fig:phlcs} shows on the left-hand side the OFE \op\nir\ light curves phased to $P_1$, and on the right-hand, the same data is phased to $P_2$.
The periodograms of the OFE light curves after removing
both sinusoidal signals are found to have peaks which are below the 0.1\% FAP line,
ensuring that the periodic signals are well fitted by sinusoids, that 
any deviation
from true sinusoids is hidden within the scatter of the data,
and that the other strong peaks in the periodograms are aliases or beats
of these two periodic signals.

When we assess in detail the significant peaks in the periodograms of the OFE light curves,
we find multiple-peaked structures due to the finite sampling of the data.  
The peaks corresponding to $P_1$ and its 
aliases, attributed to the one-day sampling of the light curves, 
are indicated in Fig. \ref{fig:periodogram} by the vertical dashed lines;
while $P_2$ and its one-day aliases are marked by the vertical dotted lines.
We also find at each significant period that there is a finely spaced three-peaked structure,
which is confirmed to arise from the seasonal (i.e., one-year) sampling of the data (see Appendix \ref{app}).  

$P_1$ is close to the orbital period of the binary ($P_{\rm orb}$ = 4.673903$\pm$0.000060~d), 
but is significantly different at a 7-$\sigma$ level. 
In order to better understand $P_1$, 
we search for periodicities in the 
residuals ($O-C$) of the EB modeling such that any period due to the EB
nature of the system would be removed from the periodograms.  We are able to again
identify both $P_1$ and $P_2$ in the $O-C$ periodograms of all observed passbands.
Table \ref{table:periods} describes in detail both identified periods in each 
observed light curve with their uncertainty, determined
via a post-mortem analysis \citep{sigmap}, for all of the OFE and $O-C$ periodograms.
We are able to verify that we have sufficient frequency 
resolution to distinguish $P_1$ from $P_{\rm orb}$ (see Appendix \ref{app}). 
Thus, we conclude that $P_1$ is not due to orbital effects, and
in particular, $P_1$ significantly differs from $P_{\rm orb}$.  
If the photometric, low-amplitude variability is caused by surface spots rotating in and out of view
on one or both of the binary components, the difference between $P_1$ and $P_{\rm orb}$ indicates
that the rotation of the stars is not fully synchronized to the orbital motion (see below). 

We measured the amplitudes of the periodic variability for both $P_1$ and $P_2$ by
simultaneously fitting two sinusoids with these periods to each light curve.
The measured amplitudes of the $P_1$ and $P_2$ signals are similar, $\sim$0.01--0.02
mag, and moreover they decrease with increasing wavelength
as expected for spot modulated variability (see Table~\ref{table:amps}).
The error of the amplitudes from the fit of the data to the double-sinusoid 
is $<$ 0.03\% in all bands.

The 4.629-d period ($P_1$) is consistent with the spectroscopically determined $v \sin{i}$ 
(17$\pm$2 and 14$\pm$3 km~s$^{-1}$ for the primary and secondary components, 
respectively; \pii) and the directly measured radii of the EB
components, $P_{\rm rot,1}/P_{\rm rot,2} = 0.88 \pm 0.22$. 
Thus we adopt $P_1$ as their rotational periods $P_{\rm rot}$.
We defer to \S\ref{sec:tides} a full discussion of $P_{\rm rot}$ in the context of
tidal evolution theory, but we note here that it is reasonable to assign the same rotation
period to both eclipsing stars.  
As the eclipsing components are being driven by tides 
toward synchronization to their orbital motion, radial contraction is changing the spin rates via conservation of angular momentum. 
In addition, \citet{zahn1989} and \citet{KhaliullinKhaliullina2011} both argue that the orbital period of \pr\ is small enough for circularization and synchronization to occur prior to the arrival on the main sequence.  
As such, the assignment of $P_1$ as the rotational period of both eclipsing components is reasonable. 
It is consistent with the independently determined observational constraints 
(i.e., $v \sin i$, $R_1$, and $R_2$), 
and moreover, it represents the conservative choice in our discussion of the tidal heating effects in \S\ref{sec:tides}.

The short period ($P_2$) is too fast 
to be due to rotation of either of the binary components; 
with measured radii of $\sim$1.7 R$_\sun$, $P_2$ would imply 
$v\sin{i} \approx 115$ km~s$^{-1}$, which is entirely inconsistent 
with the measured $v\sin{i}$ of the eclipsing components.
This periodicity, which as discussed above is 
clearly present at all epochs of our light curves spanning 15~yr,
strongly suggests the presence of a rapidly rotating third star. 
Indeed, there is ample additional evidence for the existence of 
a third star in the \pr\ system, as we now discuss.

\subsection{Characterization of a Third Stellar Component in \pr\ 
\label{thirdstar}}

In this section, we present additional evidence for a third stellar component
in the \pr\ system, which includes: (1) the presence of a featureless 
continuum in the high-resolution spectrum of \pr\ that dilutes the 
spectral features of the eclipsing components, (2) the presence
of ``third light" in the multi-band light curves which dilutes the 
eclipse depths, 
and (3) the overall spectral energy distribution of \pr, 
which is best matched by 
a third stellar photosphere plus blue excess in addition to the 
photospheres of the two eclipsing components. 
The properties of the third stellar
component are then used to refine the physical parameters that 
we determine for the EB pair in \S\ref{eb}.

\subsubsection{High-Resolution Spectroscopic Decomposition \label{disent}}

In \pii, we applied the method of tomographic decomposition on the
same multi-epoch spectra from which we determined the EB radial velocities
to recover the spectra of the individual stars, and found in that analysis
that the reconstructed spectra of the primary and secondary are 
compatible with spectral types of M1V and M3V, respectively, implying 
\teff$_{\rm ,1}$=3705~K and \teff$_{\rm ,2}$=3415~K 
\citep[from the spectral-type--\teff\ scale of][]{luh99},
which are consistent with the \teff's determined from the light curve
modeling of the system (see \S\ref{eb}).
In addition, a detailed analysis of the relative line depths of the 
reconstructed spectra made it possible to estimate their monochromatic
luminosity ratio, which was found to be $L_{\rm 1}/L_{\rm 2} \approx 1.75$
for the wavelength region around 7000\AA. This luminosity ratio 
was also shown in \pii\ to be consistent with the \teff\ ratio and 
radii ratio measured from the light curve modeling of the system.

In that analysis, we found that
the photospheric absorption lines 
appeared diluted, but we attributed this to 
poor background subtraction because the spectra used in \pii\ 
were obtained with a fiber-fed spectrograph that does not allow 
direct subtraction of the strong nebular background surrounding \pr. 
Thus here we have instead performed our analysis on the 
newly obtained high-resolution Keck/HIRES spectrum (\S\ref{hires}), which 
was obtained through a long slit permitting better background subtraction.
We extended the methods used by \citet{stem03} to the case of three spectral
components by
first constructing a model spectrum for the two eclipsing stars of \pr. 
This model spectrum is again a combination of two observed template spectra
with spectral types of $\sim$M1V and $\sim$M3V 
 (see \S\ref{hires}),
and again with a luminosity ratio of 1.75 for the region around 7000\AA\
(this luminosity ratio for the eclipsing pair from our spectral
disentangling analysis is based on the relative strengths of the 
spectral features, and thus is not a function of the
additional continuum light from the third star).
The template spectra are rotationally broadened, 
and are shifted in radial velocities, to match the widths and Doppler shifts of the 
lines  in the observed spectrum.  
The radial velocities of the template
spectra are consistent with
our final orbital solution. 
We then applied a $\chi^2$ minimization on each spectral order to
solve for any contribution of a third component.

\begin{figure}[tbp]
\includegraphics[angle=90,width=1.0\linewidth]{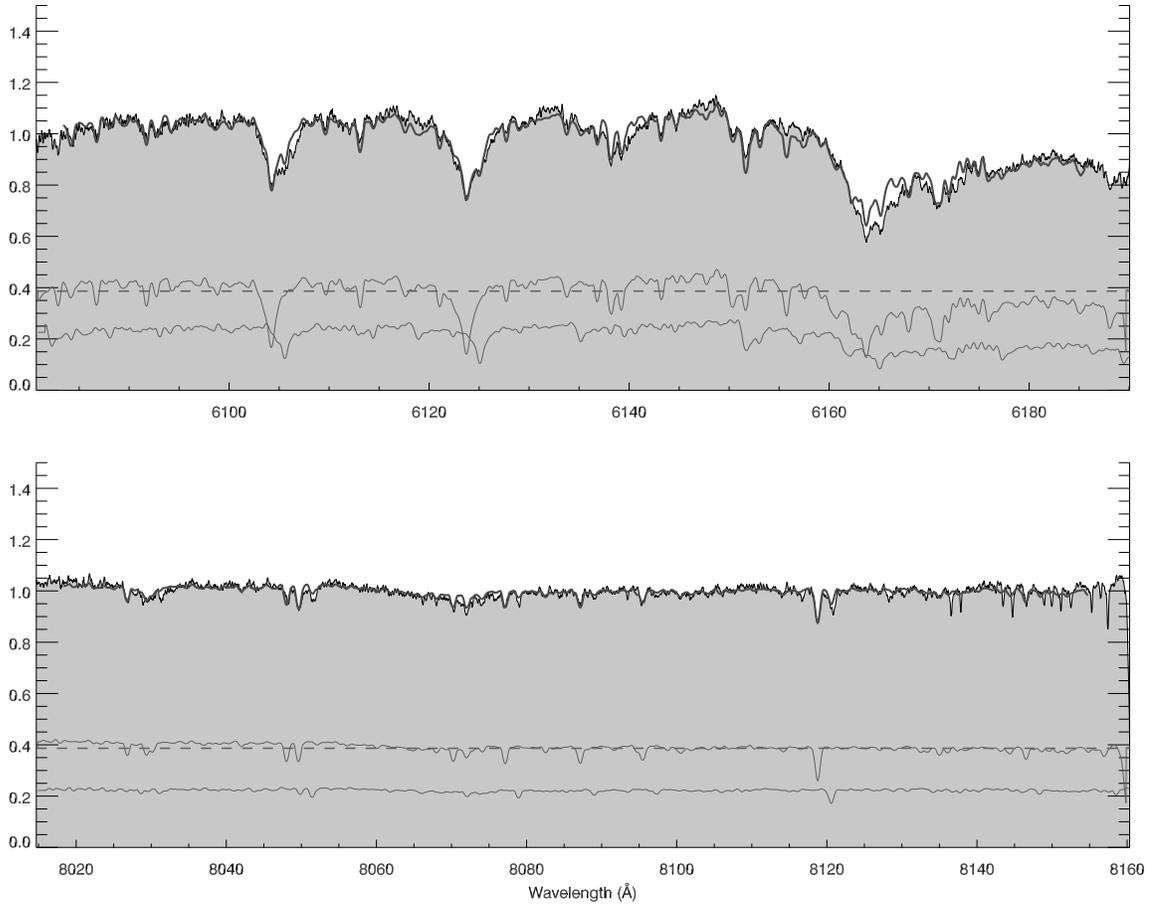}
\caption[Observed and Model Spectrum of \pr.]{\label{fig:disent}
Observed and Model Spectrum of \pr.
This figure illustrates how the observed spectrum of \pr\ 
(black solid line and gray underlying area) can be
reproduced by a simple three-component model (thick gray line). This
model consists of: an $\sim$M1V template for the primary
(upper spectrum), an $\sim$M3V template for the secondary (lower spectrum), and
a third featureless spectrum (dashed line). The components are
scaled such that the continuum ratio of the components corresponds to
0.39:0.22:0.39.  Each panel corresponds to a different order of the Keck/HIRES spectrum. 
See \S\ref{disent} for a more complete description.}
\end{figure}

We find that there is a featureless continuum present in the spectrum of \pr,
with a luminosity at 7000\AA\ that is approximately equal to that of 
the primary eclipsing component. This is illustrated for two of the 
Keck/HIRES spectral orders in Fig.~\ref{fig:disent}, 
where we show how the combined spectrum can be reproduced by adding 
the two eclipsing stellar components and a third featureless component.
The double-lined nature of the system is obvious around
the narrow absorption lines observed in the redder order shown.
The best fitting {\it normalized} luminosity ratio of all three components
is found to be (primary:secondary:extra continuum) 0.39:0.22:0.39 for the spectral 
orders shown in Fig.~\ref{fig:disent}, which correspond approximately to the 
$R_C$ and \ic\ passbands. 
The uncertainty in the normalized luminosity of the third component is 
0.15, as determined from the scatter of the measurement from the different 
spectral orders. 
 Figure~\ref{fig:disent}\  shows that one cannot reproduce the strong lines around 
6120\AA\ without additional continuum.  
Furthermore, the gravity-sensitive Ca I lines at 6103 and 6122 \AA\, present in the upper panel, 
show a good quantitative agreement in strength and shape, that could not be matched
by a different gravity and/or extra continuum.  
 This supports that one really needs the extra continuum to explain the fluxes
in \pr, and that any gravity difference between the templates and \pr\ are marginal.  
In order to further quantify this effect, we explored the effect of gravity on the 
atomic lines using synthetic spectra by decreasing $\log{g}$ from 4.5 to 3.5,
 and we find that the line depths for atomic lines increase between 0-10\%.  
This would imply that 
we are slightly overestimating the contribution of the third body, and the flux ratios would be
0.41:0.22:0.36. 
Thus, we conclude that the difference in gravity between the templates and the PMS eclipsing 
components does not affect our ability to measure the extra continuum within the quoted uncertainty. 

The analysis above does not assume anything about the nature of the
third light source. We only state that an extra featureless component is
needed in the high-resolution spectrum, and that
this is not an artifact of the reduction process. Given that 
there is no clear infrared excess in the spectral energy distribution of 
the system as would be characteristic of a disk (see \pii\ and
\S\ref{sed} below), and that the H$\alpha$ emission of several m\AA\  
seen in the eclipsing stellar components is too weak to arise from accretion (\prv),
we conclude that the third spectral component must be related to a source 
other than the two eclipsing stars.

\subsubsection{Analysis of Third Light in the \pr\ Light Curves\label{thl}}

We constrain the level of third light ($L_{3}$) in each passband 
from the spectral measurements described
above, and from the amount of third light needed to simultaneously 
fit all of the observed light curves.
The details of the EB modeling and of the exploration of the parameter 
correlations are described in \S\ref{eb}, 
as are the uncertainties of the system's fundamental parameters
introduced by the uncertainty in $L_3$. 
Here we specifically discuss $L_3$ in the context of providing additional
evidence for a third star in \pr.

The upper limit of $L_3$ allowed by the 
light curves is obtained by setting the inclination ($i$) 
of the system to 90\degr, and fitting for
$L_{3}$ as a free parameter in our modeling of the light curves
(see \S\ref{eb}).
This is the upper limit because at $i$=90\degr\ the eclipses
are intrinsically deepest, and thus the observed shallow eclipses imply
the maximum dilution. 
We find that the maximum level $L_3$ allowed by the \pr\ light curves is 
one that contributes $\sim$75\% to the total luminosity of the system in 
the \ic-band.  

To further explore the relationship between $L_{3}$ and $i$, 
we fit $L_3$ in all passbands for $i$ between 75\degr\ and 90\degr. 
We find two trends from this analysis. 
The first one is that, for any given $i$, 
the required $L_3$ is approximately constant for the \ic\nir\ light curves. 
The second trend is that $L_3$ has a {\it blue excess}, i.e., the 
$V$-band requires an {\it additional} 20\% $L_3$ contribution
to fit the eclipse depths than in the other passbands.  

Using the spectroscopic measurements described above, we are able to 
break the degeneracy between $L_3$ and $i$. 
We take $L_3$ in the \ic-band ($L_{3,I_C}$)
to be 0.39 (see \S\ref{disent}), i.e., 39\% of the system's total luminosity
($L_{{\rm tot},I_C}$).
That is, $L_{3,I_C}$ = 0.39 $\times$ ($L_{1,I_C}$ + $L_{2,I_C}$ + $L_{3,I_C}$) = 0.39 $L_{{\rm tot},I_C}$. 
Similarly, we take $L_3$ = 0.39 L$_{\rm tot}$ for the \nir-bands, since our
tests above indicated comparable $L_3$ in the \ic\nir\ light curves.
For the $V$-band, which our tests above found requires an additional 
20\% $L_3$ contribution relative to the \ic\nir\ bands, 
we therefore ascribe $L_{3,V}$ = 0.59 $L_{{\rm tot},V}$.
Even though these $L_3$ values have high uncertainties ($\sim$15\%), 
we show below that a variation in $L_3$ between 5\% and 75\% of the system's 
luminosity does not greatly affect the final physical parameters of the eclipsing components of  
\pr\ (see \S\ref{eb}).

\subsubsection{Spectral Energy Distribution of \pr\ \label{sed}}

In order to probe further into the properties of the third light source
in the system, we have attempted to model the full spectral energy
distribution (SED) of the system using NextGen model stellar atmospheres
\citep{haus99}. The SED data consist of the 12 broadband flux measurements 
described in \pii,  plus the two bluer WISE channels \citep{wise} 
covering from 0.36$\mu$m to 8.0$\mu$m. 
The WISE database has labeled the two longest WISE channels 
with the `h' flag which means 
they are likely ``ghosts" due to the very low signal-to-noise in those channels (3.8 and 2.0, respectively). 
To avoid any confusion, we have excluded the two redder WISE channels in our analysis. 
 
For each of the SED modeling attempts
described here, we held fixed the radii of the two eclipsing components,
as well as their ratio of \teff, at the values determined from the detailed
light curve modeling of the system (see \S\ref{eb}). Thus the luminosity 
ratio between the eclipsing pair is held fixed at 1.75 at 7000\AA, 
as determined from our spectral decomposition analysis (\S\ref{disent}).
We adopted a \teff\ for the primary eclipsing component of 
3675~K based on the system's reported M2 spectral type (see \S\ref{eb}).

We first attempted to model the SED by adding to the eclipsing components 
a third stellar photosphere
with \teff\ between 3000 and 6000~K,
scaled to contribute 39\% of the system's luminosity in the \ic-band (see \S\ref{thl}).
However, regardless of the \teff\ chosen for the third component, 
the $L_3$ found from our tests with the light curves (\S\ref{thl}) 
are not well reproduced by such an SED model.
For example, the {\it blue excess} (i.e., the additional 20\% 
$L_3$ in the $V$-band relative to the \ic-band; \S\ref{thl})
can be modeled by a third component with \teff$>$5000~K.
However, such a star then contributes far more third light 
in \nir\ than observed in the light curves, and moreover, the 
level of the third component's
contribution decreases with increasing wavelength. 
It is only for a third stellar component with \teff\
between 3400 and 3700~K, i.e., with a \teff\ very similar to the average \teff\
of the eclipsing components, 3560~K (see \S\ref{eb}), that the $L_3$ 
contribution remains constant at 39\% across the \ic\nir-bands. 
However, in this case the $L_3$ is also $\sim$39\% in the $V$-band, i.e., 
the observed {\it blue excess} is not reproduced.
Evidently, the source of third light cannot be a simple bare star.

\begin{figure}[tbp]
\begin{center}
\includegraphics[scale=0.6,angle=90]{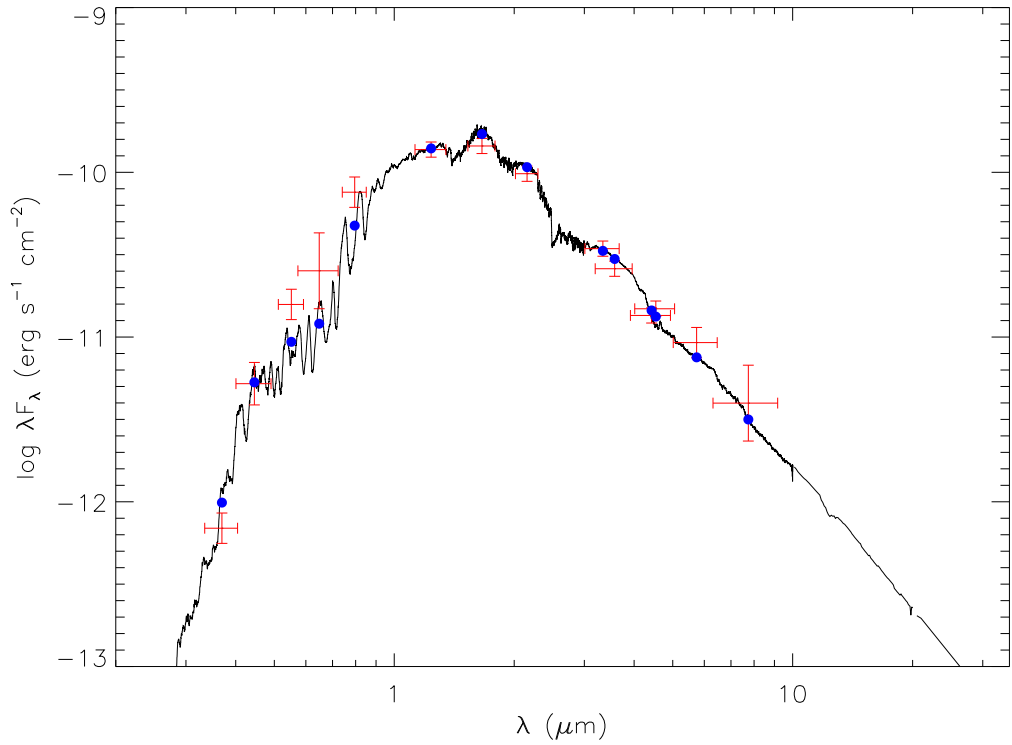}
\end{center}
\caption{\label{sedfig1}
SED fit of \pr\ including the measured \teff\ and radii of the eclipsing
pair, as well as a third star with \teff\ and luminosity equal to the
primary eclipsing star. The third star also includes a hot spot with
$T$=7500~K covering 0.1\% of the star's surface. The free parameters
of the fit are $A_V$ and distance, for which we derive 
1.2$\pm$0.6 and 440$\pm$45 pc, respectively.  
 The reduced $\chi^2$ of the fit
is 1.94. See \S\ref{sed} for details.}
\end{figure}

Finally, we again performed an SED fit 
in which we included a third stellar component,
this time fixing its temperature to the average \teff\ of the primary and secondary
eclipsing components, and once again scaling its luminosity so that it contributes 
39\% of the total system flux at \ic-band.
We also included a fourth component with a fixed \teff\ of 7500~K in
order to simulate a ``hot spot" as observed in many classical T Tauri stars
\citep[e.g.,][]{whit03}. The luminosity of this fourth component was scaled 
so that it contributes 20\% of the total flux at $V$-band (\S\ref{thl}).
The remaining free parameters of the fit are the distance to the system and
the line-of-sight extinction to the system.
The resulting best fit ($\chi^2_\nu = $ 1.94; Fig.~\ref{sedfig1})
has a distance of 440$\pm$45 pc and an extinction $A_V=$ 1.2$\pm$0.6.
These values are in good agreement with the accepted distance to the ONC
\citep[436$\pm$20 pc;][]{odell08}
and the typical extinction measured to ONC members \citep{hill97}.

We have not done a more extensive fitting of possible system parameters, 
such as different possible temperatures or filling factors for the modeled 
hot component. 
Rather, we present this SED as a plausibility check on the inferred levels
of third light measured spectroscopically and from the light curves, 
and to confirm that a putative third star with hot spot does not violate
the available SED observational constraints.
In \pii\ we performed a similar SED fit 
but including only the
two eclipsing stellar components. The fit was acceptable, though a
modest excess in the infrared portion of the SED was apparent. The new
fit presented here fits the observed fluxes very well over the entire
range 0.36--8~$\mu$m.

\subsubsection{Summary: The Third Stellar Component in Par 1802\label{thb}}

We find clear observational evidence for the existence of a third 
stellar component in the \pr\ system. The principal evidence is 
three-fold. First, there is a clear modulation of the \op\nir\ light 
curves on a very short period of 0.7355~d. This periodicity manifests
itself strongly in the periodogram analysis of the light curves at
all observed epochs spanning more than 15~yr (\S\ref{period}). Based
on the measured radii and $v\sin i$ of the eclipsing components, 
we can definitively rule out that this period is due to the rotation of
either of the two eclipsing stars.
Second, a spectral disentangling analysis applied to 
our high-resolution spectrum of \pr\ 
clearly shows the presence of added continuum which dilutes the
spectra of the two eclipsing stars (\S\ref{disent}). 
Third, our simultaneous modeling of the \op\nir\ light curves of \pr\
clearly shows third light that dilutes the eclipse depths
(\S\ref{thl}, and see also \S\ref{eb}). 
The eclipse-depth analysis also clearly indicates that,
in the \ic\ and \nir\ passbands, the third light source is characterized 
by colors very similar to those of the eclipsing stars, but that in 
$V$ the third source exhibits an additional strong ``blue excess"
similar to what is observed in Classical T Tauri stars. 
In addition to these principal lines of evidence, we have shown that the
SED of \pr\ is consistent with a simple SED model comprising the two
eclipsing stars and a third star which also includes a blue ``hot spot"
(\S\ref{sed}).
A third stellar component in \pr\ was also suggested in \prv\ by 
a long-term trend identified in the residuals of the orbit solution, 
suggesting a low-mass body in a wide, eccentric orbit.

Since the ONC is in front of a very dense, optically-thick cloud, the source of 
third light cannot be a background object and is likely to be associated with
the young cluster.
The observed short-period, low-amplitude variability can only arise from 
a rapidly rotating star and cannot be attributed to either of the eclipsing
components because there is 
no evidence for such rapid rotation in their spectra.
The rapid rotation itself suggests a young star.
An active late-type star, that is contributing
40\% of the system's luminosity and is rotating with a 0.7355-d period 
can cause the observed spot modulation ($\sim$3\% in the \ic-band) if its 
intrinsic variability is $\sim$5\%, which is within the typical observed 
variability for PMS stars.  Other other low-mass stars in the ONC have 
been found to have similarly fast rotation periods \citep[e.g.,][]{sta99}.
Moreover, if this third star is rapidly rotating and contains a strong contribution from
a hot spot as our data suggest, 
this could very well produce very shallow line profiles that are not detectable
in our spectrum and may appear as the measured additional continuum 
(see \S\ref{disent} and Fig.~\ref{fig:disent}). 
As discussed in \S\ref{sed}, including a third star with properties typical
of Classical T Tauri stars allows the broadband SED of \pr\ to be well 
fit, with a distance and extinction that are consistent with the ONC.

If the third body is indeed actively accreting
as suggested by the blue excess,
then it must be at a large enough separation
from the eclipsing pair to permit it to harbor an accretion disk.
At the distance of the ONC, the third star could be separated by as much 
as $\sim$400~AU and remain spatially unresolved in the $\sim$1 arcsec
imaging of our photometric observations.

\section{Results: Orbital and Physical Parameters of the Eclipsing Binary
Stars in \pr\ \label{eb}}

We use the Wilson-Devinney (WD) based code \phoebe\ \citep{phoebe}
to do the simultaneous modeling of the EB's radial velocity 
(RV) and light curves (LCs).
The individual RV and LC datasets are weighted inversely by the square
of their r.m.s.\ relative to the model, and the weights are updated with each
fit iteration until convergence.

In all of our fits,
we adopt the orbital period $P_{\rm orb}$ determined in \S\ref{period}. 
The rotational synchronicity parameters are calculated from the 
rotation period of the eclipsing components determined in \S\ref{period}, 
$F_{1} = F_2 = P_{\rm orb}/P_{\rm rot}$ = 1.0097$\pm$0.0013. 
We also adopt $T_{\rm eff,1}$=3675$\pm$150~K for the primary star
by assuming a primary-to-secondary luminosity ratio of 1.75 (see \S\ref{disent})
and adopting a combined \teff $=3560$~K \citep{luh99}  
from \pr's combined spectral type of M2 \citep{hill97}. 
The presence of the third star in the system does not significantly
affect this average spectral type since its \teff\ is evidently similar
to that of the eclipsing pair (see \S\ref{thirdstar}).
The uncertainty in \teff$_{\rm ,1}$ is dominated by the systematic 
uncertainty in the spectral-type--\teff\ scale for low-mass PMS stars.

\subsection{Model Fits to Radial-Velocity and Light-Curve Data}

To minimize the effect of systematic correlations in the fit parameters,
we begin our analysis by doing an initial fit to {\it only} the 
RV curves from \pii, comprised of 11 measurements for the 
primary and 9 for the secondary. 
We initially set $i$=90\degr, because the RV data
provide information only about $\sin{i}$, while $i$ is derived 
from the light curves later on.  
We utilize as our initial guesses for the RV solution the best-fit values 
from \pii\ (see Table 1 in that paper) of the parameters to be refined: 
the semi-major axis ($a \sin i$),
the mass ratio ($q \equiv M_2/M_1$),
the systemic velocity ($v_\gamma$),
and the total system mass $M \sin^3 i$.
The eccentricity ($e$) and the argument of periastron ($\omega$) 
are later determined through the fit to the RV+LC data. 
These parameters and their formal uncertainties, derived conservatively 
from the covariance matrix of the fit to the RV curves alone, 
are given in Table \ref{table:params} and are marked with
a dagger ($^\dagger$).  The resulting $a \sin{i}$, $M \sin^3{i}$, $q$, 
and $v_\gamma$ remain fixed throughout the rest of our analysis.  

We next proceed to fit the parameters that depend exclusively on 
the LC data:
$i$, 
$T_{\rm eff,2}$ (via the \teff\ ratio), 
the surface potentials $\Omega_{j}$, and the luminosities, 
without minimizing for the other parameters.  
For this task, we include the previously published \ic\ light curve and 
the \op\nir\ light curves presented in this paper (\S\ref{lcs}). 
Given that the short period, low-amplitude variability is not attributed 
to the eclipsing components but to a third body in the system, 
the light curves are first rectified by removing the sinusoidal variability 
due to the 0.7355-d period. 
We do not remove the sinusoidal varation attributed to the rotation
of the eclipsing components, as this information is encoded in the model
via the $F_1$ and $F_2$ parameters (see above).

Adopting the third light levels, $L_3$, described in \S\ref{thl}, 
we are able to fit the observed eclipse depths in all bands
to our EB model. The effects of the uncertainty in $L_3$
on the binary's physical properties is minimal and is explored in detail below. 
By fitting the RV and LC data simultaneously (RV+LC), we are able to refine $e$ and $\omega$. 
We iterate both the LC and RV+LC solutions, until we reach a consistent set of parameters 
for which the reduced $\chi^2$ of the fit is close to $\chi^2_\nu = 1$. 

\begin{figure}[tbp]
\begin{center}
\includegraphics{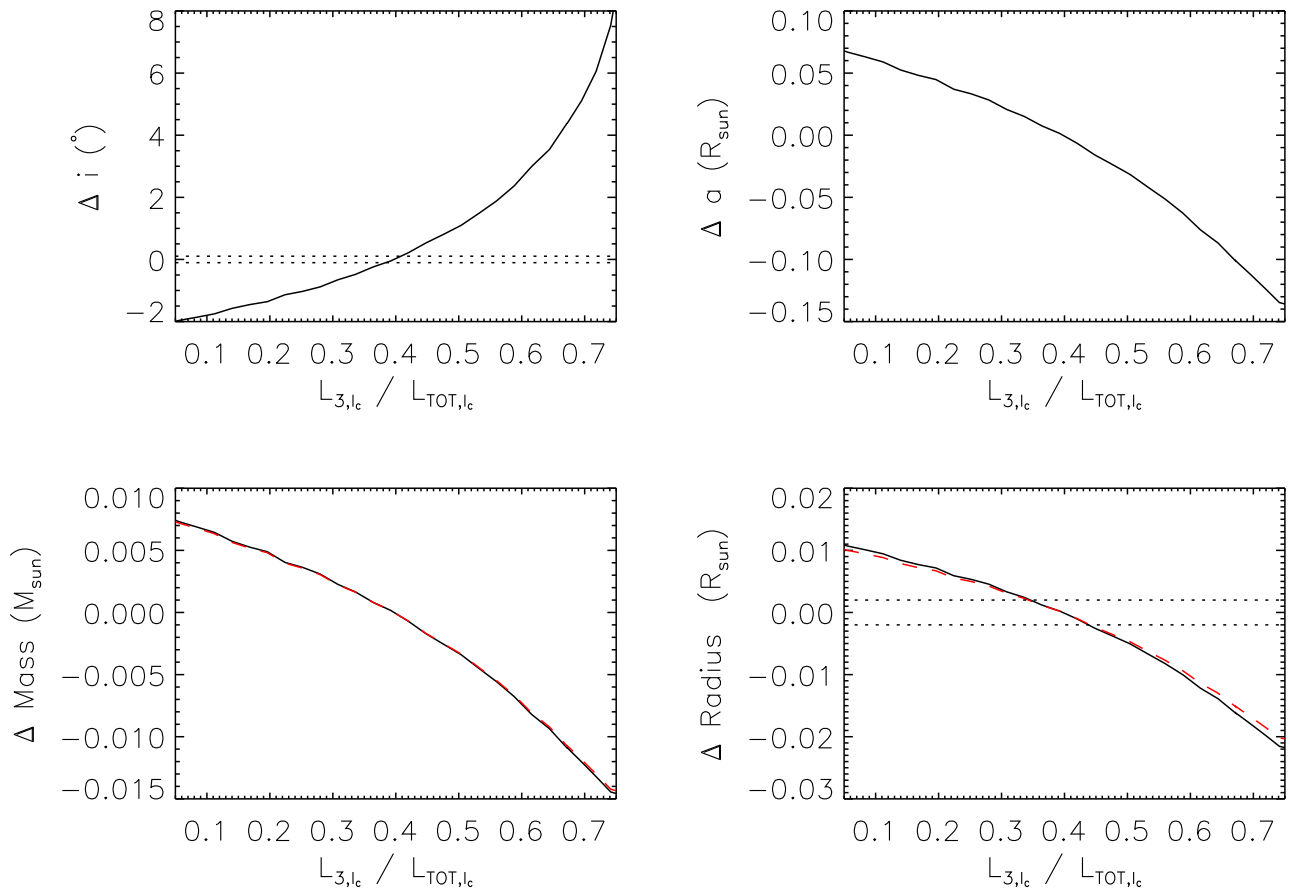}
\end{center} 
\caption[Effects of the \ic\ Third Light on the System's Parameters]{\label{fig:mrthl}
Effects of the \ic\ Third Light on the System's Parameters. 
By exploring the effects of the amount of third light on the inclination,
we are able to determine that the system's parameters, in particular those that depend
directly on $i$ (semi-major axis, radii and masses), 
do not change significantly with a change in third light.  
A variation of the level of third light in the \ic-band, 
between 5 and 75\% of the total luminosity of the system, corresponds to 
a change in inclination angle between $\sim$78 and 88\degr as shown in the top-left panel. 
The formal errors for the inclination and the radii ($\pm 1\sigma$) are 
denoted by the horizontal dotted lines in their respective panel; 
the formal uncertainties for the semi-major axis and the mass are larger than the effect
of the variation of the third light on these parameters. 
This variation of the third light, and consequently of $i$, 
corresponds to a change in the semi-major axis is of less than $\pm$ 1.5\% (top-right panel).  
It also translates into a change of less than $\pm$ 4\% in the masses, corresponding to 
less than $\pm$ 0.015 M$_\sun$ (bottom-left panel). The 
solid line and dashed line represent the change in the primary and secondary masses, respectively.
The change in the radii of the primary and secondary components of +0.01 and --0.02 R$_\sun$ 
is presented in the bottom-right panel by  
the solid line and dashed line, respectively.  
Thus, the main source of uncertainty in the determination of the inclination and the radii
is the uncertainty in the level of third light.  
}
\end{figure}

Fig.~\ref{fig:lcs} presents the observed light curves with this best-fit 
model overplotted, and the 
physical and orbital parameters of \pr\ from this model are 
summarized in Table~\ref{table:params}. 
The results from this study are generally consistent with those from \pii\ to within $\sim$1~$\sigma$ (see Table~\ref{table:params}). 
However, the system parameters are now determined more precisely, especially the eccentricity which is important for modeling the tidal evolution history of the system. 
The system inclination angle is now formally more uncertain than in \pii, but this is the result of now properly including the effects of the third light levels. However, the third light levels do not strongly affect the physical parameters (e.g., Fig.~\ref{fig:mrthl}).
The reported parameter uncertainties of our best solution include both the formal and heuristic 
parameter uncertainties (obtained from a direct $\chi^2$ mapping of the parameter space), as well as the 
uncertainties associated with our choice of third light levels, as we
now discuss.

\subsection{Effects of Third Light}

$L_3$ and $i$ are highly degenerate,
i.e., an increase in $i$ may be compensated by an increase in $L_3$,
rendering the same goodness of the fit. 
Consequently, $L_3$ most strongly impacts the parameters that depend
directly on $i$: 
$a$, the radii, and the masses.  
The \teff\ ratio is weakly dependent on a change in $i$ and its corresponding
$L_3$, because the \teff\ ratio
is constrained by the observed {\it relative depths} of the eclipses which is
itself not strongly dependent on $i$.

To explore these degeneracies as a function of $L_3$,
we vary $L_3$ in the \ic-band such that it contributes
between 5\% and 75\% of the system's total luminosity,
adjusting $L_3$ in the other bands according to the trends identified in 
\S\ref{thl}.
Fig.~\ref{fig:mrthl} shows the relationship between the change in $L_3$ and 
$i$, $a$, and the measured masses and radii of the eclipsing components. 
We find that the corresponding value for $i$ for this variation in $L_3$
lies in the range 78--88\degr.  
Since this change in $i$ is greater than its formal error of $\sim$0.1\degr, 
we adopt $\sigma_i = ^{+8.0}_{-2.0}$ degrees.
The change in the value of $a$ as $L_3$ is varied is less than 2\%. 
Thus, the masses vary by less than 4\% or 0.015 $M_\sun$.  
These changes are well below our formal uncertainty of 0.032 $M_\sun$,
which includes the above uncertainty in $i$.
The radii change by $^{+0.01}_{-0.02}$ $R_\sun$, or $\pm$1\%. 
Without including the uncertainty in $L_3$, the formal errors 
from the RV+LC fit are 0.002 $R_\sun$, for both the primary and secondary.
The main source of uncertainty in the determination of the radii is therefore the uncertainty
in $L_3$. Therefore we adopt conservatively a 1-$\sigma$ error of 0.02 $R_\sun$ 
for the radii of both eclipsing components.

\subsection{Non-Zero Orbital Eccentricity\label{sec:nonzero}}

Interestingly, our best-fit solution yields an orbital eccentricity 
that is significantly different from zero:
$e = 0.0166^{+0.0017}_{-0.0026}$.
Small eccentricities can arise spuriously because of the positive-definite
nature of $e$. Thus it was of concern that
the best-fit argument of periastron is very close to $\frac{3\pi}{2}$.
Moreover, $e$ and $\omega$ are correlated parameters through 
$e\cos \omega$ and $e\sin \omega$.
Therefore we have explored $e$ and $\omega$ in depth using two approaches.

First, we estimated $e$ and $\omega$ from simple arguments involving the
phases of primary and secondary eclipse minima, 
$t_p$ and $t_s$, and from the phase duration of each 
eclipse, $\Theta_p$ and $\Theta_s$.
The derived $e$ and $\omega$ are then related as follows \citep{kall09}: 
$e\cos{\omega} \approx \pi (t_s - t_p - \frac{1}{2}) / (1+\csc^2{i})$ and
$e \sin{\omega} \approx (\Theta_{p} - \Theta_{s})/(\Theta_{p} + \Theta_{s})$.
A lower limit for $e$ may thus be estimated by assuming $|\cos{\omega}| = 1$. 
In order to measure the separation and duration of the eclipses, 
we fit a Gaussian to both minima in the 
phased \ic-band and obtain from the phases at which they occur that their 
separation is
$t_s - t_p = 0.49799 \pm 0.00025$, 
 where the uncertainty is from the formal uncertainty on the centroids of the fitted Gaussians. 
Note that by fitting the eclipses in the entire phased light curve we are effectively 
averaging over the random scatter in the individual eclipse times (see Section \ref{period}). 
The phase separation of the eclipses differs from 0.5 by 8-$\sigma$. 
Hence, we can set as a firm lower limit, $e \geq 0.0031$.
The separation of the minima in conjunction with the measured durations, 
$\Theta_p$ = 0.1010$\pm$0.0007 and $\Theta_s$ = 0.0877$\pm$0.0012, 
render $\omega \approx 1.514\pm0.004$ $\pi$ radians. 
These $e$ and $\omega$ are estimates only, 
but demostrate that $e >$ 0 from simple theoretical 
arguments unrelated to light curve modeling.

\begin{figure}[tbp]
\includegraphics{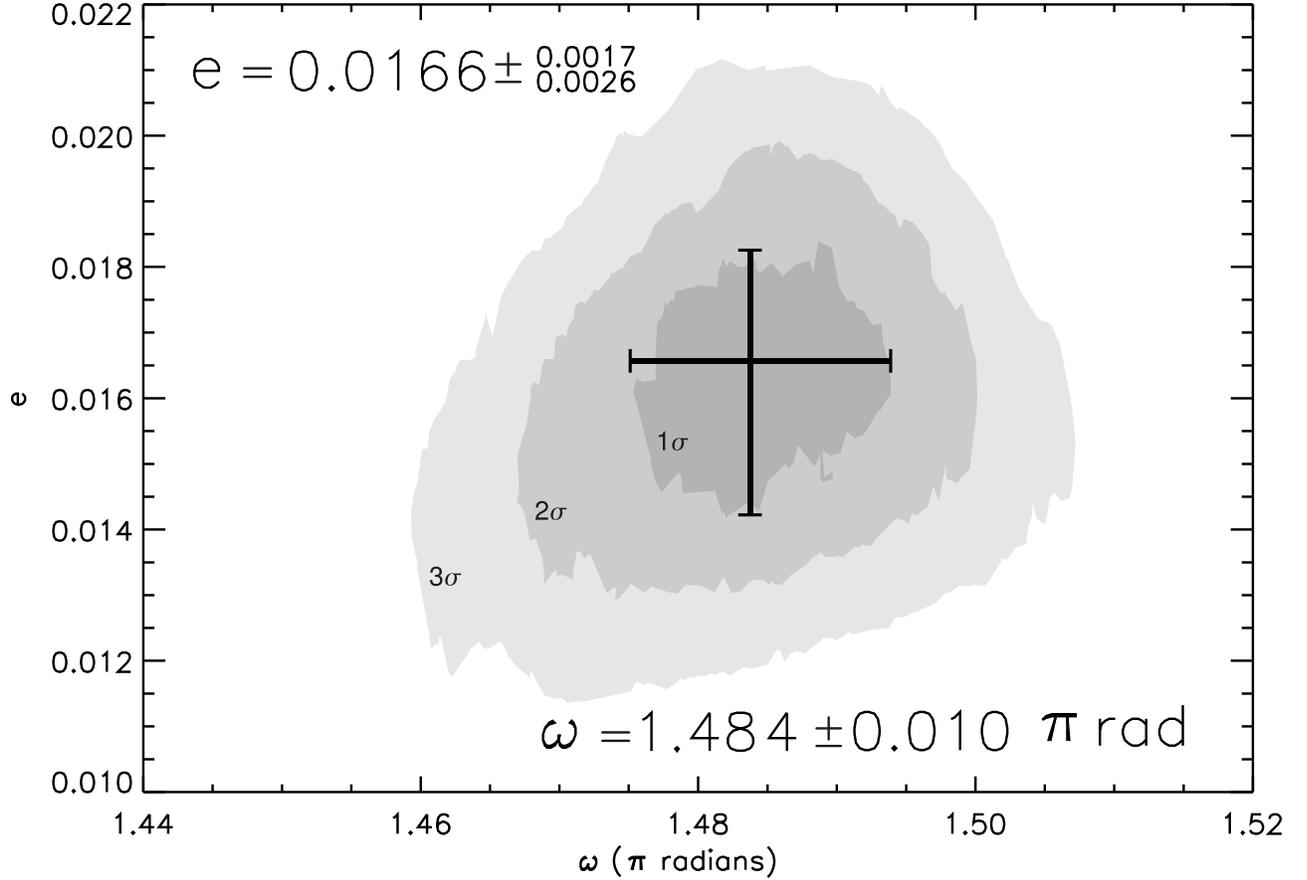}
\caption[RV+LC Joint Confidence Levels for $e - \omega$.]{\label{fig:ew}
RV+LC Joint Confidence Levels for $e - \omega$. 
Given our dataset, we are able to measure the very small but significant orbital eccentricity of the EB. 
The heuristic errors of the eccentricity $e$ and the argument of periastron $\omega$ are estimated
by the variation of a $\chi^2$-distribution with two degrees of freedom with $e$ and $\omega$.  
The center of the cross marks the point at which the $\chi^2$ of the RV+LC fit attains its minimum value;
its length and width indicate the 1-$\sigma$ uncertainties for the sampled parameters as given
by the innermost contour level.  Each subsequent contour represents a 1-$\sigma$ increase. 
The RV+LC parameter hyperspace is sampled for 0.0 $< \omega <$ 2$\pi$ and 0.0 $< e <$ 0.1;
this is the same parameter range sampled for the LC contours shown in Fig. \ref{fig:lcew}. 
}
\end{figure}

\begin{figure}[tbp]
\includegraphics{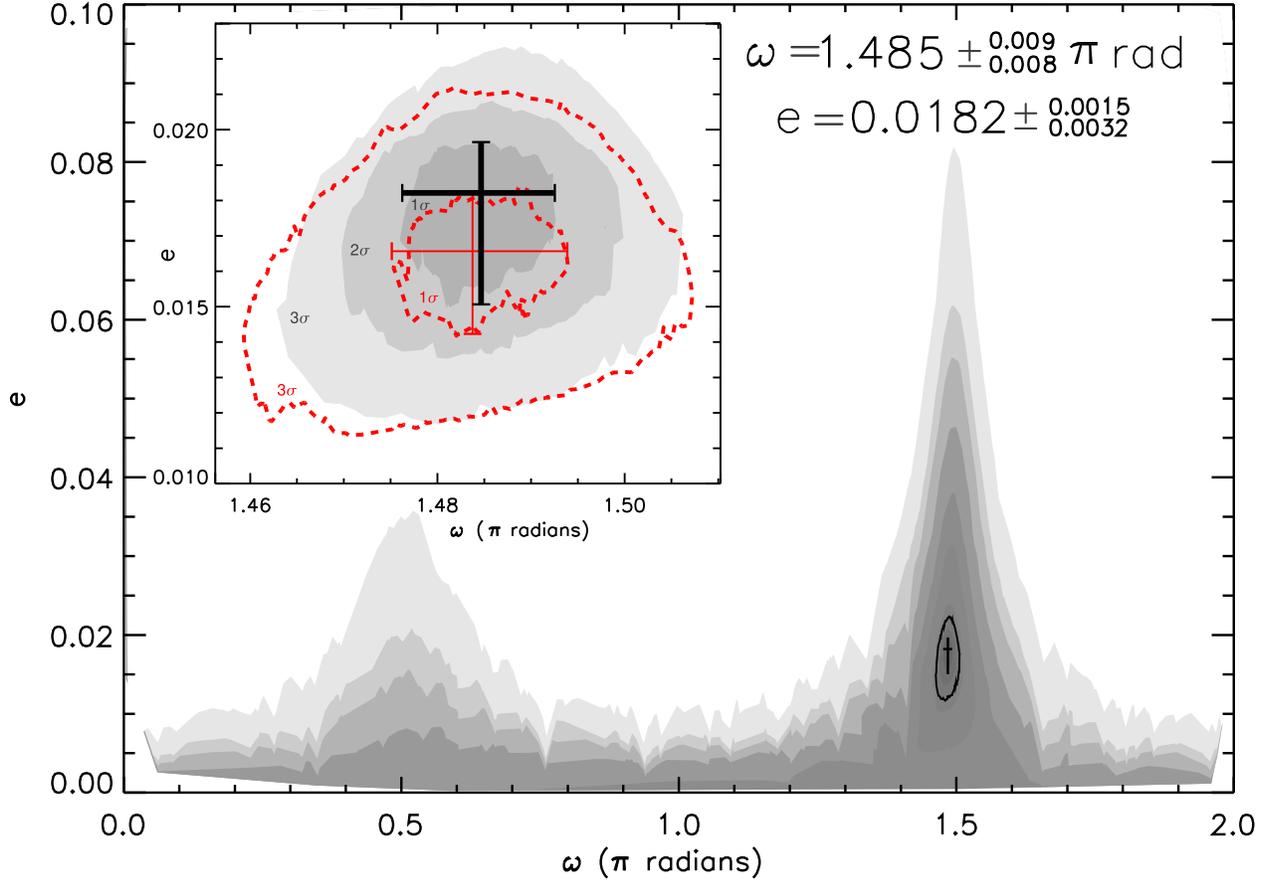}
\caption[LC Joint Confidence Levels for $e - \omega$.]{\label{fig:lcew}
LC Joint Confidence Levels for $e - \omega$.
The LC confidence contours allows us to confirm that the values for $e$ and $\omega$ from the RV+LC contours
are not systematically skewed by the weighting of the data, 
due to the abundant number of photometric data in comparison to the 
number of RV measurements.  
The figure shows the sampled parameter cross section in its entirety.
The cross marks the lowest-$\chi^2$ point to the LC fit with 1-$\sigma$ uncertainties,
 surrounded by the solid line 3-$\sigma$ confidence level.
The shaded contours beyond 3-$\sigma$ do not correspond
to a particular uncertainty level but are shown to display the two valleys in $\chi^2$ 
when the orbit's semi-major axis is parallel to the line-of-sight.  
The inset shows in detail the confidence interval for $e$ and $\omega$ within 
3-$\sigma$; and for
comparison, the dashed lines denote the 1 and 3-$\sigma$ RV+LC contours from Fig. \ref{fig:ew}.
}
\end{figure}

Second, we performed a detailed sampling of the parameter cross-section 
between $e$ and $\omega$  by fitting all of the RV and light curve data in order to determine the 
best-fit values of these parameters and their heuristic uncertainties from
a detailed examination of the shape of the $\chi^2$ space.
Fig.~\ref{fig:ew} shows the joint confidence levels for $e$ and $\omega$ 
following the variation of a $\chi^2$-distribution with two degrees of 
freedom around the RV+LC solution's minimum.
This cross section was sampled 1750 times by randomly
selecting values for $e$ in the range 0.0--0.1, and for $\omega$ in
the range 0--2$\pi$ radians. The phase shift, 
which gives the orbital phase at which the primary eclipse occurs, 
is strongly correlated with both explored parameters and is therefore 
minimized for each set of randomly selected values;
whereas the rest of the parameters are less correlated and kept constant 
at their best-fit values.    
In order to verify that $e$, $\omega$, and their uncertainties are not
artificially skewed by the weighting of both the RV and light curves 
as undertaken in \phoebe\ by WD, given that our data set is comprised 
mostly of photometric measurements, we sampled the same range in $e$ and 
$\omega$ 1900 times by fitting to the light curves alone
and obtaining their LC confidence contour levels.  
We find that the LC contours, shown in Fig.~\ref{fig:lcew}, are very 
similar to the RV+LC contours (Fig.~\ref{fig:ew}).  
The minimum value of $\chi^2$ to the RV+LC fit is 
$e$ = 0.0166$_{-0.0026}^{+0.0017}$ and $\omega$ = 1.484$\pm$0.010 
$\pi$ radians. For the LC fit, it is
$e$ = 0.0182$_{-0.0032}^{+0.0015}$ and $\omega$ = 1.485$_{-0.008}^{+0.009}$ 
$\pi$ radians. 
The detailed LC contours up to 3-$\sigma$ are shown in the inset 
in Fig.~\ref{fig:lcew}; for comparison, the 1-$\sigma$ and 3-$\sigma$
RV+LC contours are overplotted in the dashed lines. 
The two sets of contours are consistent with one another, and thus we adopt the 
values of $e$ and $\omega$ and their heuristic uncertainties from the RV+LC 
contours.

Extensive numerical integrations, like those performed for the system $\upsilon$ Andromedae \citep{barnes2011}, 
spanning the plausible range of orbits and masses of a 
third body that produce the measured eclipse timing variations (see \S\ref{period}) and 
small eccentricity 
are beyond the scope of this paper, but could be the best way to 
constrain the mass and orbit of an unseen companion.

\subsection{Temperature Ratio and Stellar radii}

\begin{figure}[tbp]
\includegraphics{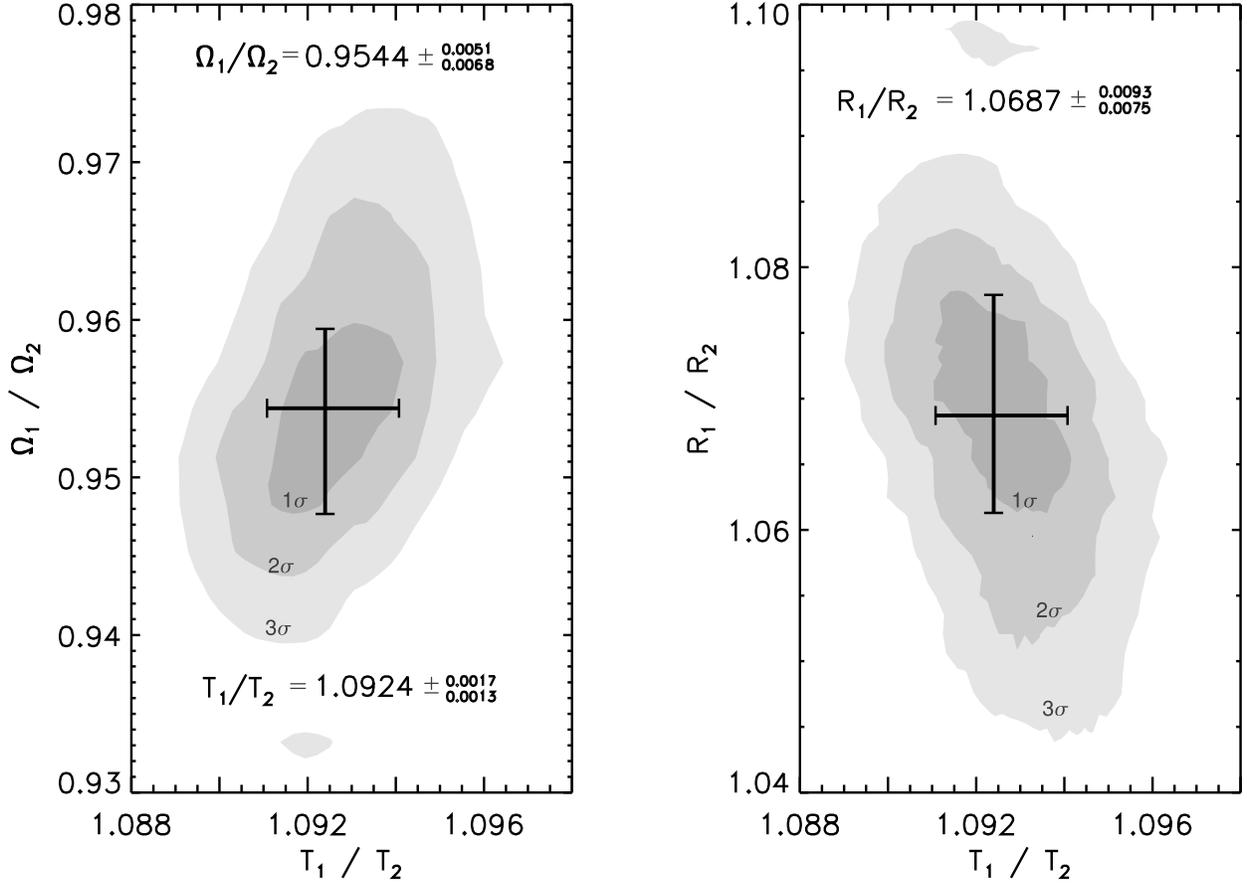}
\caption[Joint Confidence Levels for ($T_{\rm eff,1} / T_{\rm eff,2}$) -- ($R_1 / R_2$).]{\label{fig:tr}
Joint Confidence Levels for ($T_{\rm eff,1} / T_{\rm eff,2}$) -- ($R_1 / R_2$). 
Similar to Fig. \ref{fig:ew} and Fig. \ref{fig:lcew}, the significance levels given by the contours are
representative of the change in $\chi^2$ as the ratios of temperatures and radii are 
explored.  Even though the masses of the components are almost equal, $q$ = 0.985 $\pm$ 0.029,
the effective temperatures differ by 9.2$\pm$0.2\%, and the radii of the eclipsing
binary components by  6.9$\pm$0.8\%.  Consequently, their luminosities differ by 62$\pm$3\%.  
}
\end{figure}

We sampled the parameter hyperspace between 
($T_{\rm eff,1}/T_{\rm eff,2}$) and ($R_{1}/R_{2}$) 
over 2000 times, shown in Fig.~\ref{fig:tr},
in order to confirm the significance of the differences in radii and \teff\
between the eclipsing components of \pr.
We explore the \teff\ ratio in the range 1.0382--1.1271. 
The radius for the component of a detached EB
depends on the surface potentials as $\sim$$1/\Omega_j$; so the
ratio of the radii was sampled by choosing values for 
$\Omega_1$ in the range 5.5--8.4, and minimizing for $\Omega_2$. 
To facilitate the convergence of $\Omega_2$, we exploit the fact that the 
sum of the radii must remain the same due to the observational constraint
provided by the eclipse durations.
We confirm that the ratio of \teff\ as shown in \pii\ is different from unity, 
$T_{\rm eff,1}/T_{\rm eff,2}$ = 1.0924$^{+0.0017}_{-0.0013}$.
We also confirm this disparity in the case of the ratio of the eclipsing 
components radii, $R_1/R_2$ = 1.0687$^{+0.0093}_{-0.0075}$.

\section{Discussion: Possible Origins of the `Dissimilar Identical Twins'
in \pr\ \label{diss}}

\pr\ is a unique system providing important observational 
constraints in the low-mass regime at the earliest evolutionary stages.
Not only does it provide precise and direct measurements 
of the mass and the radius of each of its components; 
but because the component masses are very nearly equal 
($q=0.985\pm 0.029$, Table~\ref{table:params}), \pr\ affords a 
unique opportunity to examine the degree to which two otherwise identical 
stars in a close binary share identical evolutionary histories.
Despite having equal masses, the stars'
radii that we have measured accurately to $\sim$1\%,
differ by 7\%.  The measured \teff\ ratio, accurate to $\sim$0.2\%, 
indicates that the individual stellar \teff\ differ by 9\%.

\begin{figure}[tbp]
\includegraphics[width=0.9\textwidth]{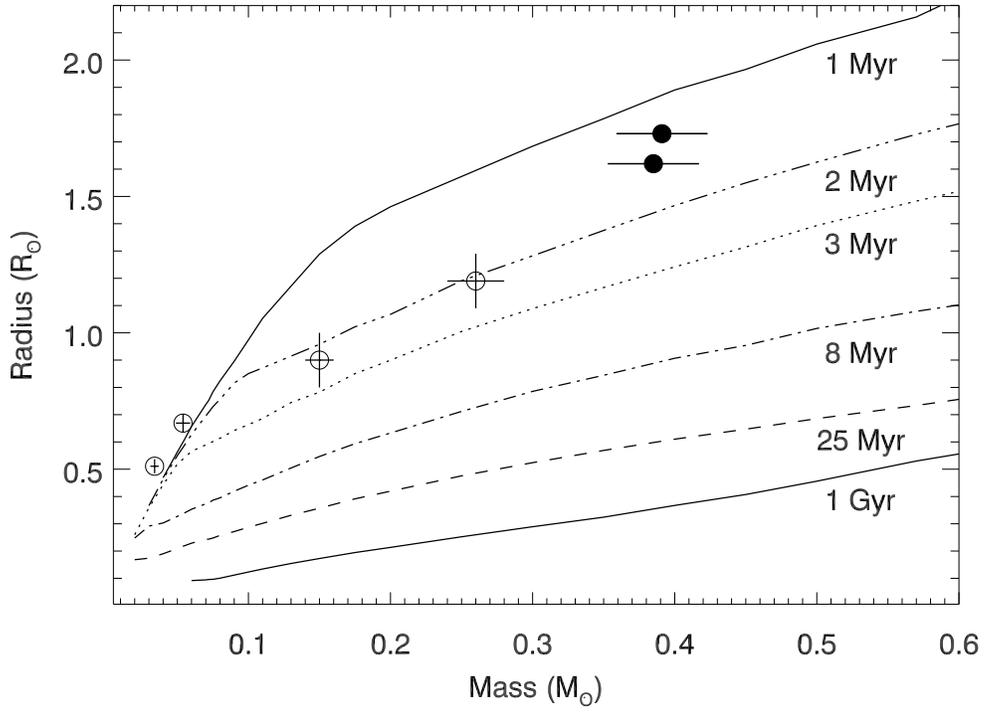}
\caption[Mass--Radius Diagram.]{Mass--Radius Diagram.
We show the mass--radius diagram comparing the measured physical properties of the youngest
and lowest mass EBs in the ONC to the 
BCAH98 theoretical isochrones with ages between 1~Myr and 1~Gyr. 
The components of \pr\ are marked by the filled circles;
JW~380 and \sm\ are shown with the
open circles.   
}\label{fig:tmr}
\end{figure}

\begin{figure}[tbp]
\includegraphics[width=0.9\textwidth]{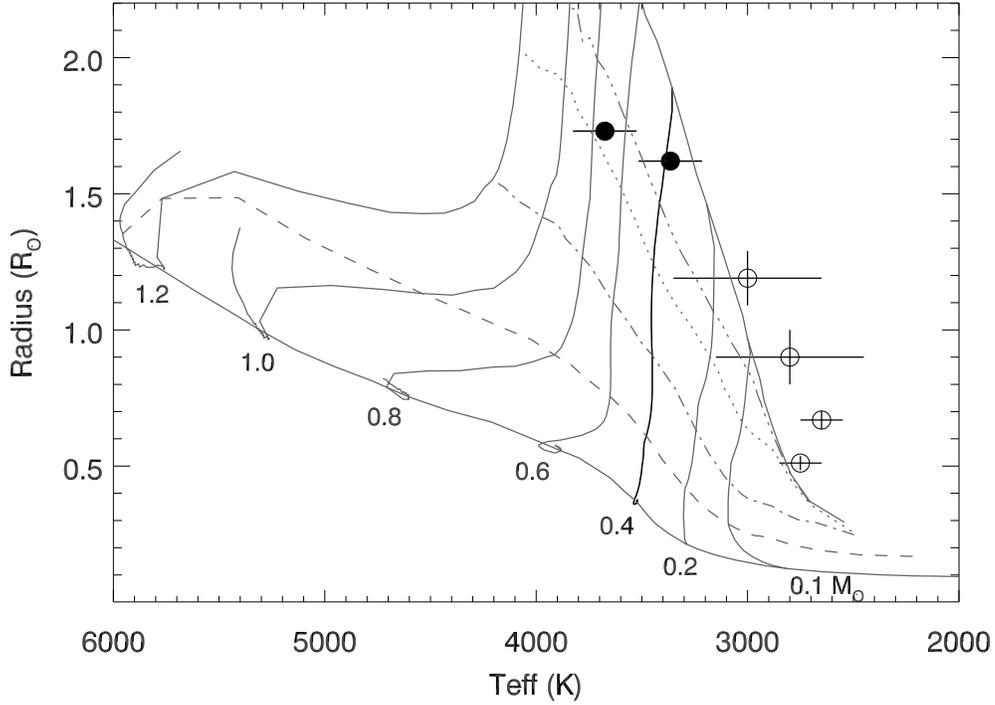}
\caption[\teff--Radius Diagram.]{\teff--Radius Diagram.
The observed EB properties and the BCAH98 theoretical isochrones 
(from 1~Myr to 1~Gyr) are denoted as described in Fig.~\ref{fig:tmr}.   
The lines that start at the 1~Myr isochrone (i.e., the solid line at far-right top corner) and descend almost vertically during
the first few Myr represent the evolution of stars of different masses (from 0.1 to 1.2 M$_\sun$) as predicted
by BCAH98.  In this diagram, 
it is clear that the components of \pr\ are young; 
however, the measured \teff\ and radii of both eclipsing components cannot be described by the 0.4 M$_\sun$ mass track.
It is only the secondary \teff\ that is consistent with the measured mass.  Additional heating mechanisms that might explain
the observed primary \teff\ are explored in \S\ref{diss}. 
}\label{fig:ttr}
\end{figure}

In this section, we consider possible implications of these physical 
differences between the two eclipsing stars in \pr.
We compare the measured properties of \pr\ to four different
pre-main sequence stellar evolutionary models:
DAM97 \citep{dant}; SDF00 \citep{SDF00}; BCAH98 \citep{bar98}, and PS99 \citep{PS99}.
As an example, in Fig.~\ref{fig:tmr}, we show
the predicted masses and radii of stars from $0.01-0.6$~M$_{\odot}$ and 
with a range of ages from 1~Myr to 1~Gyr from the BCAH98 evolutionary models
compared to the observed properties of \pr. 
In Figs.~\ref{fig:tmr}--\ref{fig:ttr}, the physical properties of the 
two other known PMS EBs in the ONC with the lowest masses and 
the youngest ages (\sm, and JW~380) are shown to provide context. 
We show these particular models because they are specifically designed to predict the properties of 
very low mass objects (late-type stars and brown dwarfs) at very young ages ($\tau \ge 1$~Myr), and they
are reasonably successful at reproducing the structural properties of these particular systems.

Despite the complex phenomena that young objects can potentially experience in their very
early evolution (i.e., accretion, magnetic activity, contraction, rapid rotation, 
tidal interactions, etc.), the observed radii of these objects 
are surprisingly well predicted by theoretical isochrones  
with an age consistent with the ONC (1--2~Myr). 
The radii of the equal-mass eclipsing components of \pr\
are enlarged, as expected for pre-main sequence stars.
However, when we look in more in detail, the measured radii of the two eclipsing
components are significantly different, 
and this is not predicted by a single theoretical isochrone. 
Moreover, the effective temperatures of these two stars are 
also significantly different.  In the temperature-radius diagram that compares the BCAH98 models with 
the observed properties from the PMS EBs (Fig.~\ref{fig:ttr}), this implies
that the two equal-mass stars cannot both be fit by the same mass track. 

In the first 10~Myr, as these low mass stars descend along
the Hyashi track to the main sequence, a rapid contraction in radius 
at roughly constant temperature is predicted.  All the models we examined show similar trends from 1--10~Myr,
however the BCAH98 models predict a cooler temperature for this contraction by $\sim$200~K than the 
other three models.  Furthermore, the DAM97 model is unique in that it shows an additional rapid 
evolution in \teff\ prior to the first 1 Myr (as shown in \pii).  
Despite some genuine successes, no existing single star evolutionary model 
(that does not include accretion, magnetic activity, tidal heating, rotation, detailed convection)
is able to reproduce the observed properties of both eclipsing components of \pr\ with a single age and mass. 

Moreover, the different predictions by each of the theoretical models 
lead to different possible physical interpretations for \pr.  
The \teff\ and radius of the secondary star is well reproduced by the BCAH98 models for a 0.4~\msun\ 
star with an
age of ~1--2 Myr (see Fig.~\ref{fig:ttr}), but the primary star is too hot for its mass.  However, the models by DAM97, PS99, and SDF00 
predict a 2 Myr, 0.4~\msun\ star to have a \teff\ consistent with the primary star, but 
overestimate the temperature of the secondary.  This comparison 
suggests that one of the two components (probably the primary star) 
may have experienced some form of additional heating making it unexpectedly hotter than its twin.
In addition, as discussed in \pii, the DMA97 models suggest a small age difference could be invoked to explain
the differences in physical properties between the two eclipsing components of \pr\ if the system is hotter by $\sim$250~K and younger than 1 Myr (see \S\ref{dis:age}).  
We consider the possible explanations below in more detail.

\subsection{Magnetic Activity}

Evolutionary models have typically not included the effects of magnetic fields 
because of the complexity and difficulty involved in their modeling.
However, the effects of magnetic fields are thought to be the cause of the enlarged radii and cool temperatures of field M dwarfs
found in eclipsing binary systems \citep{lop07}.  
The presence of spots and/or the reduction of the convective efficiency of the star, due to increased magnetic activity, lower the
effective temperature and increase the radius in order to maintain the star's luminosity \citep{cha07}.
\pr's nearly equal-mass components 
should have similar convection zone depths and are rotating at similar rates, thus
it is likely that they have similar magnetic activity levels.  Moreover, the measured 
H$\alpha$ emission of both stars is weak.  
If magnetic activity were the cause of the discrepant radii and effective temperatures in \pr, 
we would expect the
cooler component to have the larger radius.  However, we find the opposite. 
The secondary star has the smaller radius and
cooler temperature, thus magnetic activity is unlikely to be causing the disparate
radii and temperature reversal
found between the  twin components of \pr.

\subsection{Competitive Accretion}\label{dis:age}
As discussed in \pii, a difference in age of a few $\times 10^5$~yr 
could potentially explain the observed differences in
\teff\ and radius for the eclipsing stars in \pr.
The idea here is that mass equalizing mechanisms during the binary 
formation process may have preferentially directed accretion 
from the circumbinary disk to the
(initially) lower-mass component, leading that star to cease the phase
of heavy accretion later than its companion, and causing its ``birth" to
be effectively delayed relative to its companion 
(i.e., causing it to appear younger).
This ``competitive accretion" scenario has been specifically
advanced in the context of \pr\ by \citet{simon09}.

If the \teff's for both stars could be shifted $\sim$250~K
hotter (while preserving the accurately measured \teff\ ratio),  
implying a $\sim$2$\sigma$ shift relative to the likely systematic
uncertainty on the absolute \teff\ scale for these stars, 
the stars are best matched by the DAM97 models, 
which predict that a 0.4~M$_\sun$ star {\it decreases} in \teff\ during 
the first Myr. This would imply that the primary star is 
the younger component, being both hotter and larger.
In this scenario, the primary will presumably evolve along the 
0.4~M$_\sun$ track and, within a few Myr, appear identical to its
(presumed slightly older) twin.

\subsection{Tidal Evolution and Heating \label{sec:tides}}

Another potential explanation for the observed differences in luminosity
of the \pr\ EB components is the presence of additional energy sources.
We have determined that the orbit of \pr\ is not circular,
but rather has a non-zero eccentricity of $e$=0.0166$\pm$0.003.
In addition, we have measured the rotation period of the EB components to be
very close to but significantly different than the orbital period
(\S\ref{period}).
Consequently, the EB components should be experiencing some degree of
tidal interaction.
In this section, we consider the role of tides and the amount of tidal
heating that the two stars may have experienced during their lifetimes
in order to reproduce their observed physical properties.
In particular, we wish to determine whether the primary star could have
acquired enough additional tidal heating to explain its apparent
over-luminosity relative to its twin.

A substantial body of research is devoted to tidal theory.
The reader is referred to
\citet{Hut1981}, \citet{FerrazMello2008}, \citet{Leconte2010},
 \citet{Mazeh2008}, \cite{Zahn2008}  
and references therein for a more complete description of the
derivations and nuances of various theoretical treatments. For this
investigation, we consider the so-called ``constant phase lag" (CPL)
and ``constant time lag" (CTL) models, the details of which are
provided in Appendix~\ref{app-tides} \citep[and see][]{Heller2011}.
Our approach is not intended to be a
definitive treatment of the tidal evolution of this binary. Rather we
 estimate tidal effects using standard assumptions and linear
theory. More detailed modeling could prove enlightening, but is beyond
the scope of this investigation.  Even so, the discussion below
indicates that standard assumptions suggest tidal heating is important in this binary.

The CPL and CTL models assume that the physical properties of the stars
are constant with time. However, as the \pr\ system is very
young ($\sim$1~Myr), the radii are expected to be contracting quickly.
This contraction could have a profound effect on
tidal processes as the radius enters both the CPL and the CTL models
at the fifth power (see Eqs.~\ref{eq:Zp} and \ref{eq:Z}).
Radial contraction will also enter into the angular momentum evolution through
the rotational frequency (Eqs.~\ref{eq:o_cpl} and \ref{eq:o_ctl}).
Thus we have added radial contraction to the CPL and CTL models,
in a manner similar to that in \citet{KhaliullinKhaliullina2011}, but
note that their treatment does not include obliquity effects.
\cite{dant} and \cite{bar98} provide from their calculations the
time rate of change of the radius, $dR/dt$ in R$_\odot$/Myr, for 0.4~M$_\odot$ stars.
We fit their models with a third order polynomial using Levenberg-Marquardt
minimization,

\begin{equation}\label{eq:drdt}
\frac{dR}{dt} = a_0 + a_1t + a_2t^2 + a_3t^3,
\end{equation}

\noindent where $a_0 ... a_3$ are constants listed in Table~\ref{tab:rory}.
For simplicity, we assume the radial contraction is independent of the
tidal evolution. Therefore, the ``radius of gyration'' $r_g$, i.e., the moment of inertia is $M(r_gR)^2$,
is held constant, and moreover, we can
express the change in stellar spin due solely to radial contraction as

\begin{equation}
\label{eq:spinrate}
\frac{d\omega}{dt} = -\frac{2\omega}{R}\frac{dR}{dt}.
\end{equation}

\begin{figure}[tbp]
\includegraphics{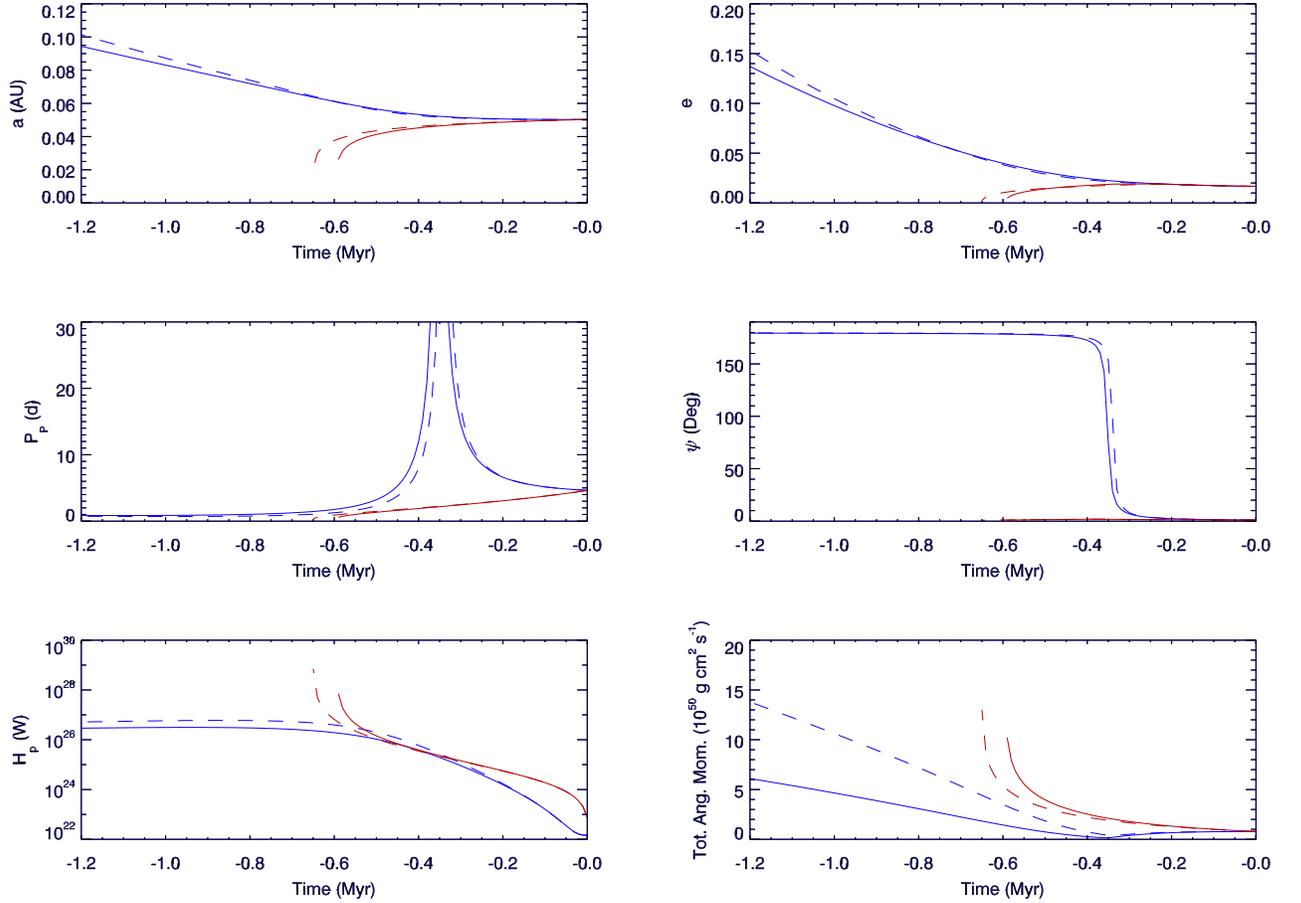}

\caption{\label{fig:radcontr}History of the eclipsing components of \pr\ due to tidal
  processes \textit{and} radial contraction for $\psi_P = 1^\circ$ and
  $\psi_S = 0$. Blue curves are for the CTL model, and red the
  CPL. Solid curves use the \cite{bar98} radial contraction
  model, dashed the \cite{dant} model. 
  \textit{Top Left}: Semi-major axis. \textit{Top Right}:
  Orbital eccentricity. \textit{Middle Left}: Primary's rotation
  period. \textit{Middle Right}: Primary's obliquity. \textit{Bottom
  Left}: Primary's tidal heat. \textit{Bottom Right}: The total
  angular momentum of the system. 
}

\end{figure}

In Fig.~\ref{fig:radcontr}, we present the history of \pr\ due to both
tidal evolution (for both the CPL and CTL models) and radial contraction
(for both the DM97 and BCAH98 stellar evolution models).
The behavior of the resulting tidal evolution history is qualitatively
different as compared to the evolution without radial contraction effects
(see Appendix~\ref{app-tides}).
We assume the primary star's obliquity $\psi_P$ = 1\degr\ at the present time, i.e., $t=0$,
in order to be able to determine the evolution of $\psi_P$.
The CTL model evolution (blue curves) predicts that the stellar
obliquities were anti-parallel up to 0.5~Myr ago, and then
 $\psi_P$ rapidly ``flipped" (actually its rotation was halted and
then reversed).
For the CPL model (red curves), the evolution breaks down at $\sim$0.5~Myr
in the past as the stars are predicted to have been merged
(and consequently the model fails to conserve angular momentum within
a factor of 10; bottom right panel).
In principle this could be taken as a constraint on the system's maximum
age, but more likely this reflects the limitations of the simplified
linear tidal theory that we have adopted; the model is  unable to
account for effects such as Roche lobe overflow that would certainly
have been important if the stars had once been in physical contact.
This behavior of the CPL model could be avoided by tuning the model's
$Q$ parameter  (here we have adopted the standard value for
pre-main sequence stars of $10^6$ \citep[see][]{Zahn2008}),
however for the current discussion we discard the
prediction of formerly merged stars and instead adopt the CTL model
predictions as more physically plausible.

\begin{figure}[tbp]
\includegraphics{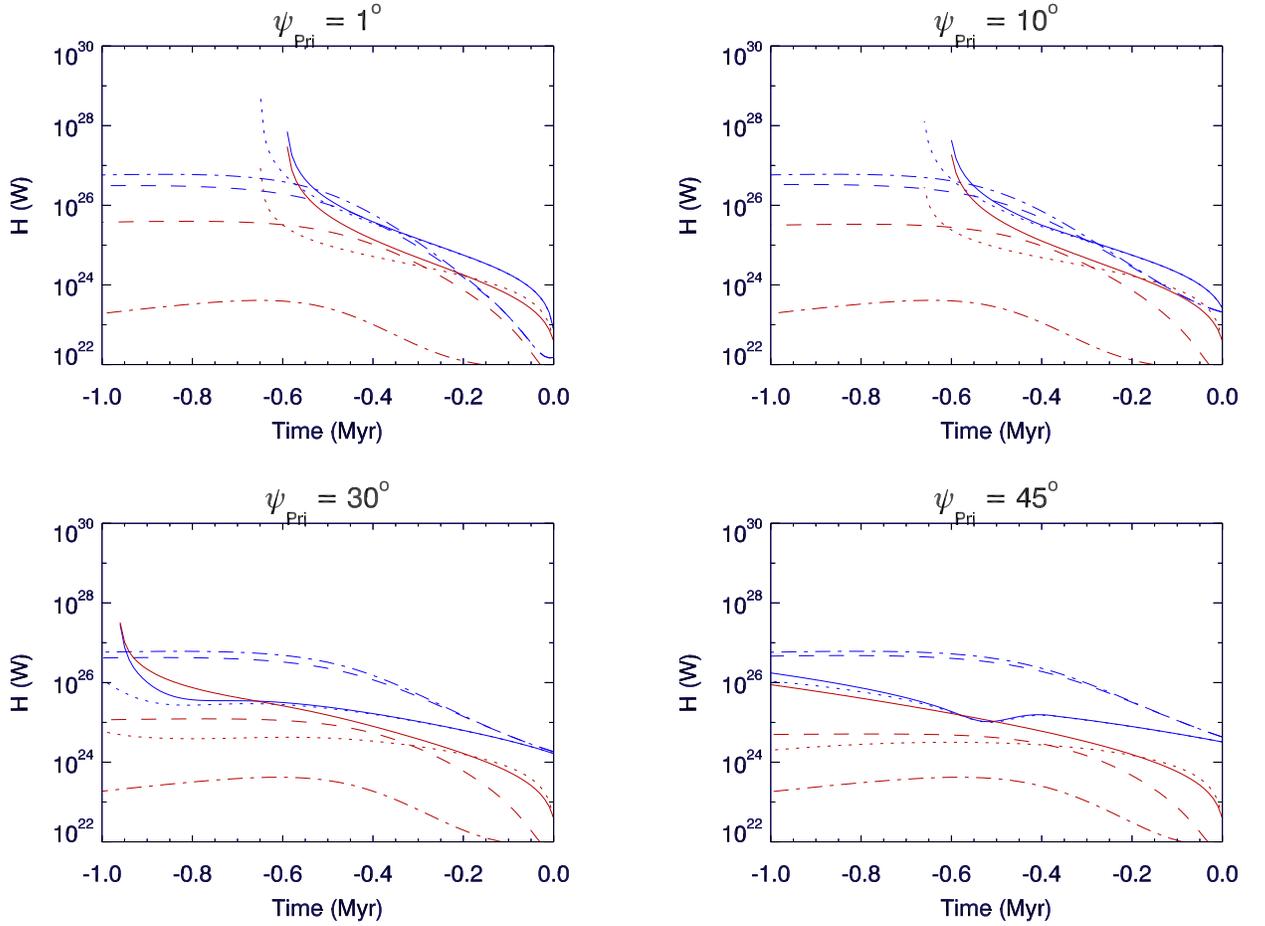}

\caption{\label{fig:heat}Comparison of heating histories of the two eclipsing 
components of \pr. The primary is represented by blue curves, secondary
by red. In all cases the obliquity of the secondary is zero. Solid curves
assumed the CPL and \cite{bar98} models, dotted CPL and \cite{dant}
models, dashed CTL and \cite{bar98}, and the dash-dotted the CTL
and \cite{dant} models. From top left to bottom right}, the current obliquity
of the primary is $1^\circ, 10^\circ, 30^\circ$, and $45^\circ$.

\end{figure}

Finally, in Fig.~\ref{fig:heat} we consider whether tidal effects can
explain the increased radius of the primary. We consider a range of
possibilities, all of which assume $\psi_S = 0$ for simplicity. There are
four possible combinations of models: CTL-B98, CPL-B98, CTL-DM97, and
CPL-DM97. The CPL models (solid and dotted curves) do not predict
much difference in the tidal heating rates between the primary and
secondary star. The CTL-B98
model (dashed curve) predicts about a factor of 10 difference between
the stars, and with the primary receiving $>10^{26}$~W as recently as
0.2~Myr ago ($\psi_P = 45^\circ$). The model that predicts the
largest difference in the heating rates between the two stars is
the CTL-DM97 case (dot-dashed curves), which predicts a difference in
heating of more than 3 orders of magnitude as recently as 0.5~Myr
ago. Moreover, the heating rate seems to level out at about
$10^{26}$~W (and roughly independent of $\psi_P$).
However, the CTL models, as shown in Fig.~\ref{fig:radcontr},
predict that the obliquities were anti-parallel when the heating rates
are most different.

A tidal heating rate of $\sim10^{26}$~W is comparable to the observed
difference between the \pr\ primary and secondary stars' luminosities
($\Delta L \approx 7\times 10^{25}$~W; see Table~\ref{table:params}).
The models in Fig.~\ref{fig:heat} are able to supply at least this
amount of tidal heating to the primary up to $\sim$0.2~Myr in the past.
Thus, assuming that the primary star will have retained this extra heat for
at least 0.2~Myr,
we conclude that tidal processes are able to provide a sufficient
energy source to explain the differences in the luminosities of the
\pr\ EB pair. However, this requires the spins to have been misaligned,
at least in the past. Observation of misaligned spin axes in \pr\ (e.g.,
via the Rossiter-McLaughlin effect) would provide strong additional
evidence in favor of this interpretation, 
though it may be a challenging observation if the obliquity is small.

\section{Summary \label{summary}}

Parenago 1802 is a pre--main-sequence, double-lined, detached 
eclipsing binary, 
and is the youngest known example of a low-mass
system with a mass ratio of unity ($q$=0.99$\pm$0.03).
It presents a unique source of  observational constraints for 
low-mass stars during the early stages of their evolution.   
Contrary to what theoretical evolutionary models predict 
for stars of the same mass, composition and age, 
the radii of the eclipsing pair differ by 6.9$\pm$0.8\%, 
their effective temperatures differ by 9.2$\pm$0.2\%,
and consequently their luminosities differ by 62$\pm$3\%,
despite their masses being equal to within 3\%.

The \pr\ system appears to include an unresolved, low-mass third star 
that is rapidly rotating and likely accreting (i.e., a Classical
T~Tauri star) in a wide orbit about the eclipsing pair.
This third star manifests itself in multiple ways, including 
a very short period modulation of the light curves, 
excess continuum in the spectra, and 
dilution of the eclipse depths in the light curves.
The broadband spectral energy distribution of the \pr\ system can
be modeled by two stars with the measured properties of the eclipsing
pair, plus a third low-mass star including an accretion hot spot, at
a distance of 440$\pm$45~pc, consistent with the distance to the Orion
Nebula Cluster.

We measure the rotation period of the eclipsing stars
and find that they are rotating with a period that is slightly but
significantly faster than the orbital period.
Moreover, the orbit has not yet circularized, presenting
a small but significant eccentricity, $e$=0.0166$\pm$0.003.
These orbital and rotational characteristics provide
important insight into the tidal interactions at a very young age that 
lead to the synchronization and circularization of binaries.
We show that tidal interactions during the past 1~Myr history of the
system could plausibly have injected sufficient heat into one of the
eclipsing components to explain its over-luminosity relative to its twin.
This explanation predicts that the epoch of high tidal heating terminated
$\sim$0.2~Myr ago, and thus requires that the over-luminous component 
have retained the tidal heat for at least the past 0.2~Myr (assuming
a nominal system age of 1~Myr). This explanation also
predicts that the eclipsing pair's rotation axes may yet be misaligned, 
which could be observable via the Rossiter-McLaughlin effect.

\acknowledgments
We gratefully acknowledge NSF funding support through grants 
AST-0349075 and AST-1009810.
Some of the data presented herein were obtained at the W.M.\ Keck
Observatory, which is operated as a scientific partnership among the
California Institute of Technology, the University of California and the
National Aeronautics and Space Administration. The Observatory was made
possible by the generous financial support of the W.M.\ Keck Foundation.
The authors wish to recognize and acknowledge the very significant cultural
role and reverence that the summit of Mauna Kea has always had within
the indigenous Hawaiian community.  We are most fortunate to have the
opportunity to conduct observations from this mountain.

\appendix

\section{Verification of Periodicity Analysis with Synthetic Signals\label{app}}

\begin{figure}[tbp]
\includegraphics{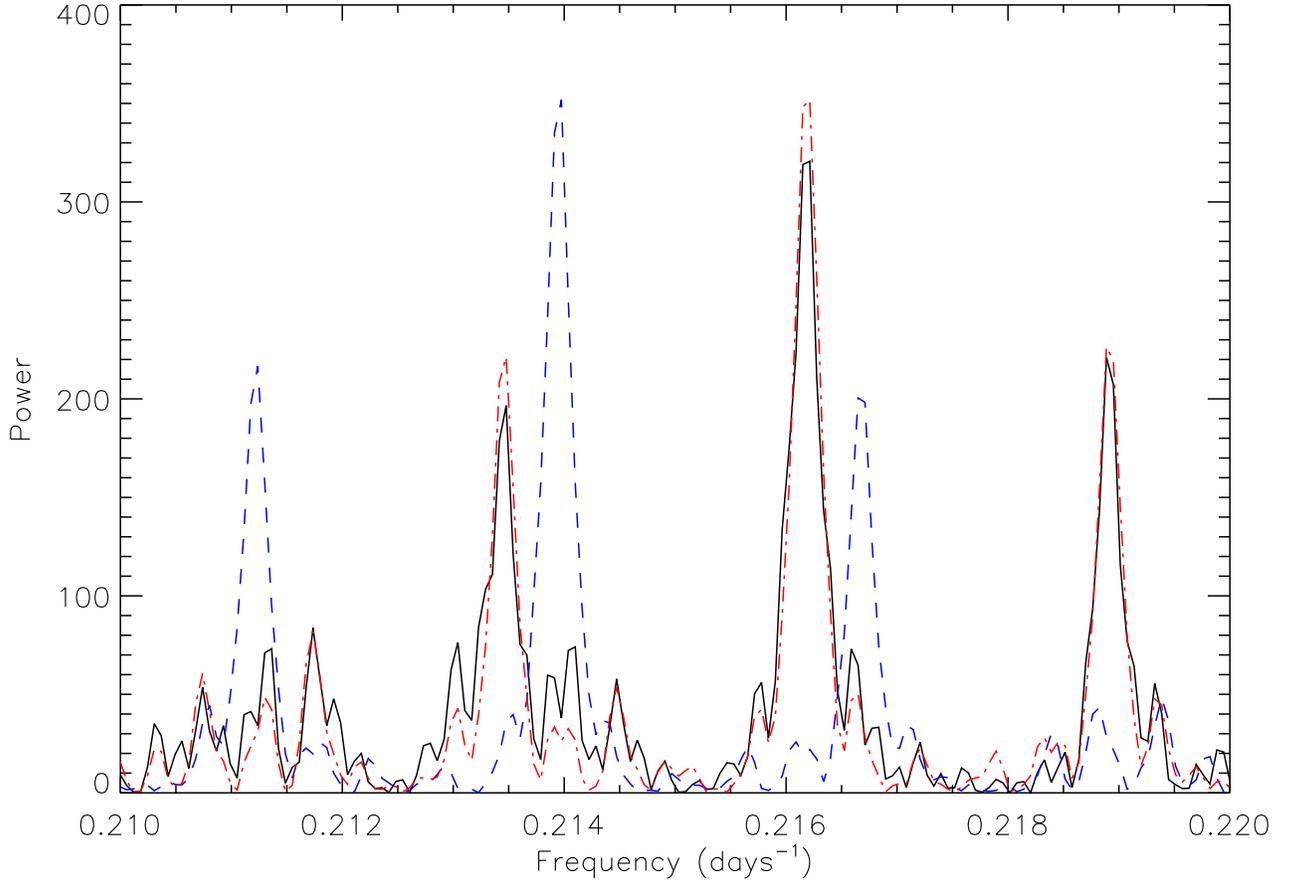}
\caption[OFE \ic\ and Synthetic Periodograms.]{\label{fig:synper}
OFE \ic\ and Synthetic Periodograms.
We compare the periodogram of the OFE \ic\ light curve (solid line)
around the frequency of 1/$P_1$ $\simeq$ 0.216 d$^{-1}$ with two synthetic sinusoidal signals,
 one with a period equal to $P_1$ (dashed-dotted line)
and another with a period of $P_{\rm orb}$ (dashed line).
Both synthetic signals have been sampled to the timestamps of the \ic\ data
to preserve its statistical characteristics; and their periodograms
have been scaled to the amplitude of the OFE \ic\ periodogram.
The three-peaked structure around the most significant peak is due to the yearly sampling
of the light curve;
the side-peaks are separated from the central peak by a frequency of 1/360 d$^{-1}$.
Since we are able to clearly distinguish between the periodogram peaks
of the $P_1$ signal and those of the $P_{\rm orb}$ signal,
we conclude the $P_1$ is significantly distinct from the orbital period.
}
\end{figure}

To ensure that the peak that corresponds to $P_1$, identified as the rotation 
period of the eclipsing components, is significantly
different than that of the orbital period given the available dataset,
we create two synthetic sinusoidal signals that are sampled using the
timestamps of the OFE \ic\ light curve: one with a period equal to $P_1$
and another to $P_{\rm orb}$.  After running the synthetic signals through
the periodicity analysis described in \S\ref{period}, we compare their
periodograms to that of the OFE \ic\ light curve.  Fig.~\ref{fig:synper}
shows that the periodogram of the OFE \ic\ light curve (solid line)
around the frequency of 1/$P_1$ $\simeq$ 0.216 d$^{-1}$ is almost equal
to the normalized periodogram of the synthetic signal with the same period
(Fig.~\ref{fig:synper}, dash-dotted line), as expected.  Moreover,
the periodogram of the synthetic signal with a period equal to the orbital
period (Fig.~\ref{fig:synper}, dashed line) is clearly distinct from
the other two periodograms.  By directly assessing the {\it window function}
of the data through the periodograms of the synthetic periodic signals,
we are able to discard the possibility that the three-peaked structure
found in the periodograms centered around the most prominent peaks is an
artifact of our periodicity analysis.  The periodograms of the synthetic
signals, as shown in Fig. \ref{fig:synper}, also present the three-peaked
structure confirming that it arises from the sampling of the data and that
we have enough frequency resolution to discern $P_1$ from $P_{\rm orb}$.

\section{Details of Tidal Evolution Models\label{app-tides}}

For our calculations of tidal evolution, we employ ``equilibrium
tide'' models, originally derived by \citet{Darwin1879,Darwin1880}. 
These models assume the
gravitational potential of the tide raiser can be expressed as the sum
of Legendre polynomials and that the elongated equilibrium shape of
the perturbed body is slightly misaligned with respect to the line
which connects the two centers of mass. This misalignment is due to
dissipative processes within the deformed body, i.e., friction, causing a secular 
evolution of the orbit as well as the angular momenta of the two bodies. 
When consistently calculating the tidal interaction of two bodies, the roles of the 
tide raiser and the perturbed body can be switched.
This approach leads
to a set of six coupled, non-linear differential equations, but note that
the model is in fact linear in the sense that there is no coupling
between the surface waves which sum to the equilibrium shape.

\subsection{The Constant Phase Lag Model \label{subsub:cpl}}

In the ``constant-phase-lag'' (CPL) model of tidal evolution, the angle between the line connecting the
centers of mass and the tidal bulge is assumed to be constant. This
approach has the advantage of being analogous to a damped driven
harmonic oscillator, a well-studied system, and is quite commonly
utilized in planetary studies \citep[e.g.,][]{GS1966}. In this case, the
evolution is described by the following equations

\begin{equation}\label{eq:e_cpl}
  \frac{\mathrm{d}e}{\mathrm{d}t} \ = \ - \frac{ae}{8 G M_1 M_2} \sum_{i \, \neq \, j}Z'_i \Bigg(2\varepsilon_{0,i} - \frac{49}{2}\varepsilon_{1,i} + \frac{1}{2}\varepsilon_{2,i} + 3\varepsilon_{5,i}\Bigg)
\end{equation}

\begin{eqnarray}\label{eq:a_cpl}
  \frac{\mathrm{d}a}{\mathrm{d}t} = \frac{a^2}{4 G M_1 M_2} \sum_{i \, \neq \, j} Z'_i  {\Bigg(} 4\varepsilon_{0,i} + e^2{\Big [} -20\varepsilon_{0,i} + \frac{147}{2}\varepsilon_{1,i} + \nonumber \frac{1}{2}\varepsilon_{2,i} - 3\varepsilon_{5,i} {\Big ]} \\ 
 -4\sin^2(\psi_i){\Big [}\varepsilon_{0,i}-\varepsilon_{8,i}{\Big ]}{\Bigg )} 
\end{eqnarray}

\begin{eqnarray}\label{eq:o_cpl}
  \frac{\mathrm{d}\omega_i}{\mathrm{d}t} \ = \ - \frac{Z'_i}{8 M_i r_{\mathrm{g},i}^2 R_i^2 n} {\Bigg (}4\varepsilon_{0,i} + e^2{\Big [} -20\varepsilon_{0,i} + 49\varepsilon_{1,i} + \varepsilon_{2,i} {\Big ]} \\ 
+ \nonumber \ 2\sin^2(\psi_i) {\Big [} -2\varepsilon_{0,i} + \varepsilon_{8,i} + \varepsilon_{9,i} {\Big ]} {\Bigg )}   
\end{eqnarray}

\begin{equation}\label{eq:p_cpl}
  \frac{\mathrm{d}\psi_i}{\mathrm{d}t} \ = \ \frac{Z'_i \sin(\psi_i)}{4 M_i r_{\mathrm{g},i}^2 R_i^2 n \omega_i} {\Bigg (} {\Big [} 1-\xi_i {\Big ]}\varepsilon_{0,i} + {\Big [} 1+\xi_i {\Big ]}{\Big \{}\varepsilon_{8,i}-\varepsilon_{9,i}{\Big \}} {\Bigg )} \ ,
\end{equation}

\noindent where $e$ is eccentricity, $t$ is time, $a$ is semi-major axis, $G$ is
Newton gravitational constant, $M_1$ and $M_2$ are the two masses,
$R_1$ and $R_2$ are the two radii,
$\omega$ is the rotational frequency, $\psi$ is the obliquity, 
$r_g$ is the ``radius of gyration,'' i.e., the moment of inertia is $M(r_gR)^2$,  
and $n$ is the mean motion. The quantity $Z'_i$ is

\begin{equation}\label{eq:Zp}
Z'_i \equiv 3 G^2 k_2 M_j^2 (M_i+M_j) \frac{R_i^5}{a^9} \ \frac{1}{n Q_i} \ ,
\end{equation}

\noindent where $k_2$ is the Love number of order 2, and tidal $Q$ is the ``tidal quality factor.'' The parameter $\xi_i$ is

\begin{equation}\label{eq:chi}
\xi_i \equiv \frac{r_{\mathrm{g},i}^2 R_i^2 \omega_i a n }{ G M_j},
\end{equation}

\noindent where $i$ and $j$ refer to the two bodies. 
The signs of the phase lags are

\begin{equation}\label{eq:epsilon}
\begin{array}{l}
\varepsilon_{0,i} = \Sigma(2 \omega_i - 2 n)\\
\varepsilon_{1,i} = \Sigma(2 \omega_i - 3 n)\\
\varepsilon_{2,i} = \Sigma(2 \omega_i - n)\\
\varepsilon_{5,i} = \Sigma(n)\\
\varepsilon_{8,i} = \Sigma(\omega_i - 2 n)\\
\varepsilon_{9,i} = \Sigma(\omega_i) \ ,\\
\end{array}
\end{equation}

\noindent with $\Sigma(x)$ the sign of any physical quantity $x$, thus
$\Sigma(x)~=~+~1~\vee~-~1~\vee~0$. 

The tidal heating of the $i$th body is due to the transformation of
rotational and orbital energy into frictional heating. The heating
from the orbit is 

\begin{equation}\label{eq:E_orb_cpl}
\nonumber
\dot{E}_{\mathrm{orb},i} = \ \frac{Z'_i}{8} \ \times \ {\Big (} \ 4 \varepsilon_{0,i} + e^2 {\Big [-20 \varepsilon_{0,i} + \frac{147}{2} \varepsilon_{1,i} + \frac{1}{2} \varepsilon_{2,i} - 3 \varepsilon_{5,i} {\Big ]} - 4 \sin^2(\psi_i) \ {\Big [}\varepsilon_{0,i} - \varepsilon_{8,i}{\Big ]} \ {\Big )}} \ ,
\end{equation}

\noindent and that from the rotation is 

\begin{equation}\label{eq:E_rot_cpl}
\nonumber
\dot{E}_{\mathrm{rot},i} = \ - \frac{Z'_i}{8} \frac{\omega_i}{n} \ \times \ {\Big (} \ 4 \varepsilon_{0,i} + e^2 {\Big [}-20 \varepsilon_{0,i} + 49 \varepsilon_{1,i} + \varepsilon_{2,i}{\Big ]} + 2 \sin^2(\psi_i) \ {\Big [}- 2 \varepsilon_{0,i} + \varepsilon_{8,i} + \varepsilon_{9,i}{\Big ]} 
\ {\Big )} \ .
\end{equation}

\noindent The total heat input rate into the $i$th body is therefore 

\begin{equation}\label{eq:E_tide_cpl}
\dot{E}_{\mathrm{tide},i}^{\mathrm{CPL}} = - \ (\dot{E}_{\mathrm{orb},i} + \dot{E}_{\mathrm{rot},i}) > 0 \ .
\end{equation}

It can also be shown \citep{gold66,murr99} that the  
equilibrium rotation period for both bodies is

\begin{equation}\label{eq:p_eq_cpl}
P_{\rm eq}^{CPL} = \frac{P}{1 + 9.5e^2}
\end{equation}

\noindent for low $e$ and no obliquity. However, given the discrete nature of the CPL model, we caution
that integration of Eqs.~\ref{eq:e_cpl}--\ref{eq:p_cpl} will not always yield the value predicted by
Eq.~\ref{eq:p_eq_cpl}.

In the CPL model, for a given 
$\psi$ the rate of evolution and amount of heating are 
set by three free parameters: $Q$, $k_2$, and $r_g$. 
We choose for these parameters $10^6, 0.5$, and 0.5, respectively. 
These choices are consistent with observations of other 
stars \citep{lin96,jack09}. 

\begin{figure}[tbp]
\includegraphics{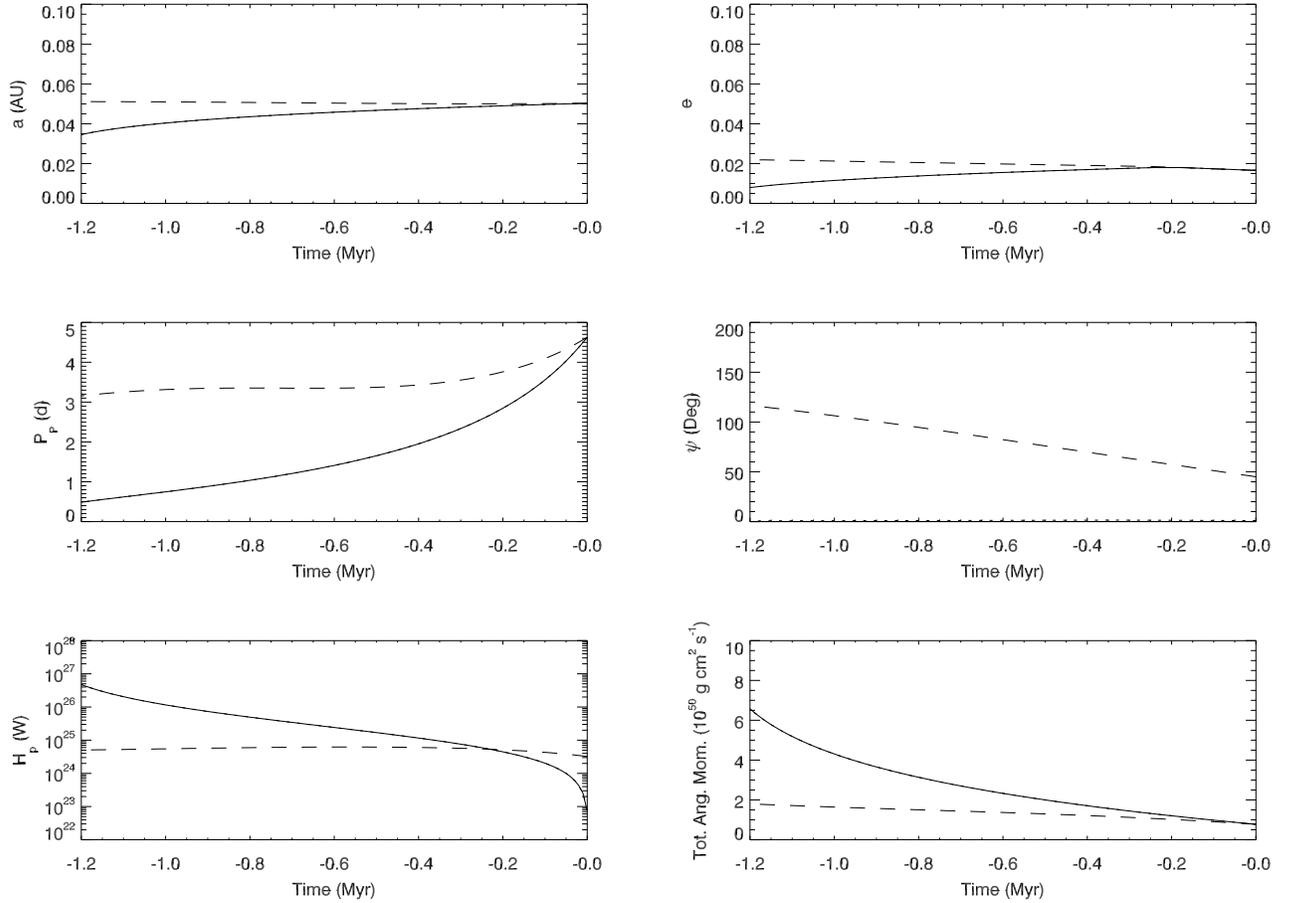}

\caption{\label{fig:cpl}
  History of \pr\ due to tidal
  processes, and assuming the constant phase lag (CPL) model, for three
  different current obliquities of the primary: 0 (solid curves),
  1$^\circ$ (dotted curves; indistinguishable from the solid curves
  in this case), and 45$^\circ$ (dashed curves). 
  The system's observed properties are adopted as the ``initial values"
  at $t=0$ and the tidal evolution equations evolved back in time. 
  For context the evolution is shown slightly beyond the system's
  nominal age of 1~Myr.
  \textit{Top Left}: Semi-major axis. \textit{Top Right}:
  Orbital eccentricity. \textit{Middle Left}: Primary's rotation
  period. \textit{Middle Right}: Primary's obliquity. \textit{Bottom
  Left}: Primary's Tidal Heat. \textit{Bottom Right}: Total
  angular momentum (orbital $+$ spin) of the system.
}

\end{figure}

To illustrate the behavior of the CPL model,
the history\footnote{We evolve the system backward in
time by adopting the currently observed system properties as the
``initial values" ($t=0$) and then solving the differential equations
for negative times, up to the system's nominal age ($t = -1$~Myr).}
of \pr\ including solely the effects of CPL evolution is shown in 
Fig.~\ref{fig:cpl} for three choices of $\psi_P$, the obliquity of the 
primary. 
Note that for low $\psi_P$ the model predicts that the eccentricity 
of the \pr\ system was smaller in the past. This may seem counterintuitive,
since in most binary systems where tidal effects are considered, the
eccentricity tends to circularize over time. However, in the scenario
where only CPL effects are considered, the stars previously rotated 
faster than the binary orbit, and thus the tidal forces induce an
acceleration of the stars in the same direction as the orbit, leading
to an increase of $e$.

As shown in Fig.~\ref{fig:cpl}, bottom right panel, the CPL model
conserves the total system angular momentum (orbit $+$ spin) in the 
calculation to within a factor of a few out to 1~Myr in the past.
Strictly speaking the calculation should conserve total angular 
momentum, because the model dissipates tidal energy not angular momentum,
and there are no angular momentum sources or sinks in the model. 
The lack of angular momentum conservation is a result of the linearization
of the model, and thus the degree to which angular momentum is not
conserved can be regarded as a measure of the degree to which the simple
assumptions of the model are breaking down.
For our purposes here, where we seek to investigate order-of-magnitude
tidal heating effects, we regard angular momentum conservation to within
a factor of a few as acceptable.

\subsection{The Constant Time Lag Model \label{subsub:ctl}}

The constant-time-lag (CTL) model assumes that the time interval between the passage of the
perturber and the tidal bulge is constant. This assumption allows the
tidal response to be continuous over a wide range of frequencies,
unlike the CPL model. However, if the phase lag is a function of the
forcing frequency, then the linear approach is not valid, as the
system is no longer analogous to a damped driven harmonic
oscillator. Therefore, this model should only be used over a narrow
range of frequencies, see \cite{Greenberg2009}. This requirement is met for \pr, except where noted.

The orbital evolution is described by the following equations

\begin{equation} \label{eq:e_ctl}
  \frac{\mathrm{d}e}{\mathrm{d}t} \ = \ \frac{11 ae}{2 G M_1 M_2} \sum_{i \, \neq \, j}Z_i \Bigg(\cos(\psi_i) \frac{f_4(e)}{\beta^{10}(e)}  \frac{\omega_i}{n} -\frac{18}{11} \frac{f_3(e)}{\beta^{13}(e)}\Bigg)
\end{equation}

\begin{equation}\label{eq:a_ctl}
  \frac{\mathrm{d}a}{\mathrm{d}t} \ = \  \frac{2 a^2}{G M_1 M_2} \sum_{i \, \neq \, j} Z_i \Bigg(\cos(\psi_i) \frac{f_2(e)}{\beta^{12}(e)} \frac{\omega_i}{n} - \frac{f_1(e)}{\beta^{15}(e)}\Bigg)
\end{equation}

\begin{equation}\label{eq:o_ctl}
  \frac{\mathrm{d}\omega_i}{\mathrm{d}t} \ = \ \frac{Z_i}{2 M_i r_{\mathrm{g},i}^2 R_i^2 n} \Bigg( 2 \cos(\psi_i) \frac{f_2(e)}{\beta^{12}(e)} - \left[ 1+\cos^2(\psi) \right] \frac{f_5(e)}{\beta^9(e)} 
\frac{\omega_i}{n} \Bigg)  
\end{equation}

\begin{equation}\label{eq:p_ctl}
  \frac{\mathrm{d}\psi_i}{\mathrm{d}t} \ = \ \frac{Z_i \sin(\psi_i)}{2 M_i r_{\mathrm{g},i}^2 R_i^2 n \omega_i}\left( \left[ \cos(\psi_i) - \frac{\xi_i}{ \beta} \right] \frac{f_5(e)}{\beta^9(e)} \frac{\omega_i}{n} - 2 \frac{f_2(e)}{\beta^{12}(e)} \right) 
\end{equation}

\noindent where

\begin{equation}\label{eq:Z}
 Z_i \equiv 3 G^2 k_{2,i} M_j^2 (M_i+M_j) \frac{R_i^5}{a^9} \ \tau_i \ ,
\end{equation}

\noindent and 

\begin{equation}\label{eq:f_e}
\begin{array}{l}
\beta(e) = \sqrt{1-e^2},\\
f_1(e) = 1 + \frac{31}{2} e^2 + \frac{255}{8} e^4 + \frac{185}{16} e^6 + \frac{25}{64} e^8,\\
f_2(e) = 1 + \frac{15}{2} e^2 + \frac{45}{8} e^4 \ \ + \frac{5}{16} e^6,\\
f_3(e) = 1 + \frac{15}{4} e^2 + \frac{15}{8} e^4 \ \ + \frac{5}{64} e^6,\\
f_4(e) = 1 + \frac{3}{2} e^2 \ \ + \frac{1}{8} e^4,\\
f_5(e) = 1 + 3 e^2 \ \ \ + \frac{3}{8} e^4.
\end{array}
\end{equation}

\noindent The tidal heating of the $i$th body is therefore

\begin{equation}\label{eq:E_tot_ctl} \nonumber 
\dot{E}_{\mathrm{tide},i}^{\mathrm{CTL}} = \ Z_i {\Bigg (} \frac{f_1(e)}{\beta^{15}(e)} - 2 \frac{f_2(e)}{\beta^{12}(e)} \cos(\psi_i) \frac{\omega_i}{n} + \ {\Big [} \frac{1+\cos^2(\psi_i)}{2} {\Big ]} \frac{f_5(e)}{\beta^9(e)} {\frac{\omega_i^2}{n^2}} {\Bigg )} \ .
\end{equation}

It can also be shown that the equilibrium rotation period for both bodies is

\begin{equation}\label{eq:p_eq_ctl}
P_{{\rm eq},i}^{CTL} = \frac{2 \pi \beta^3(e) f_5(e)}{n f_2(e)} \frac{1 + \cos^2{(\psi_i)}}{2 \cos{(\psi_i)}}.
\end{equation}

In the CTL model, $Q$ is replaced by the ``time lag,''
$\tau$. 
For the limiting case of $e=0$ and $\psi=0\degr$ the two parameters are related as $Q = 1/n\tau$ \citep{Leconte2010,Heller2011}. 
For \pr, $n = 1.56 \times 10^{-5}$, corresponding to $\tau = 0.064$~s at $t=0$.
We therefore choose this time lag, and as a result the CTL model
predicts about the same rate of change as the CPL model near $t=0$,
but note that the CPL and CTL evolutions diverge as the orbital period 
changes ($n = 2\pi/P$) into the past.

\begin{figure}[tbp]
\includegraphics{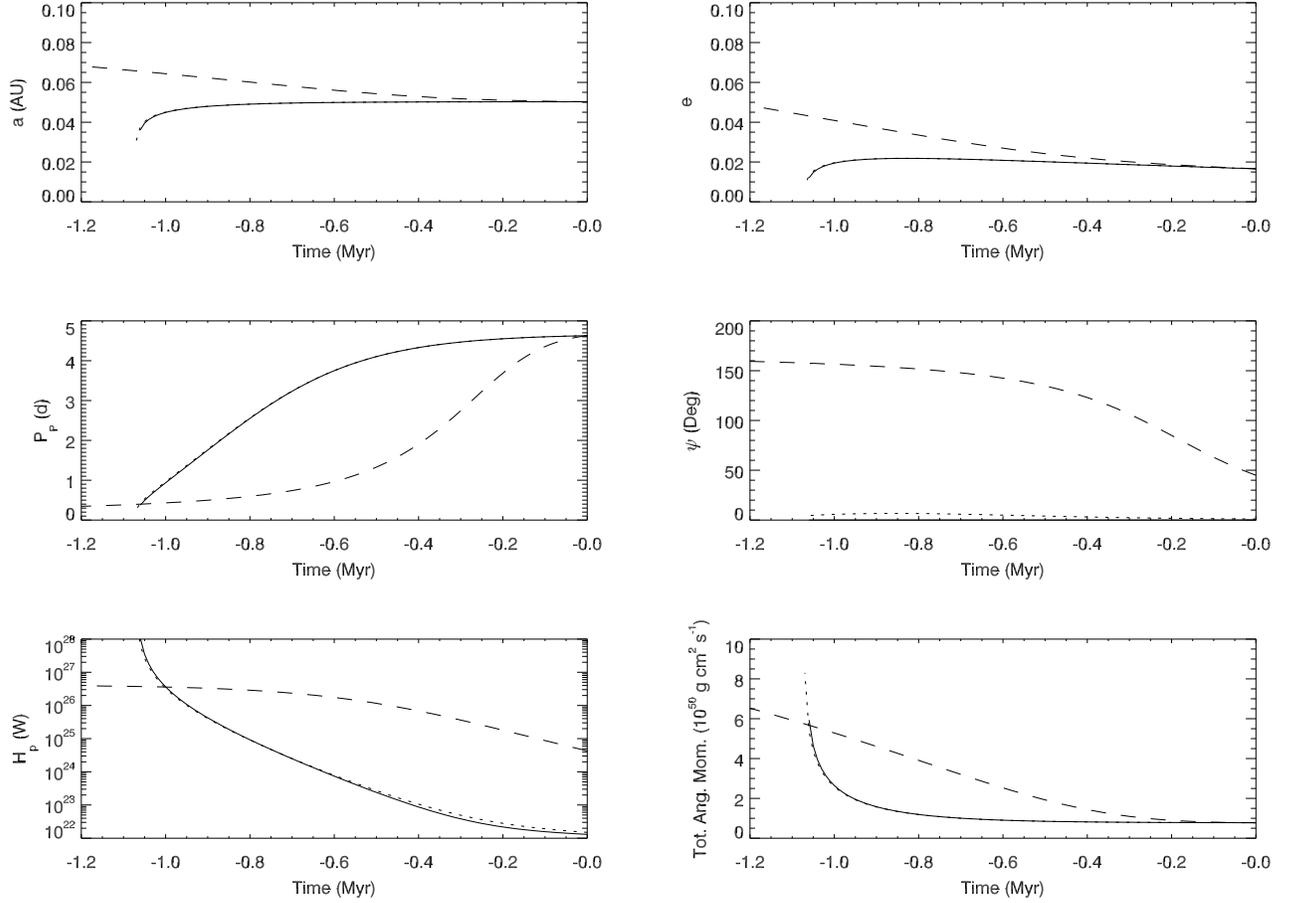}

\caption{\label{fig:ctl}History of \pr\ due to tidal
  processes, and assuming the constant-time-lag model (CTL), for three
  different current obliquities of the primary: 0 (solid curves),
  1$^\circ$ (dotted curves) and 45$^\circ$ (dashed
  curves). \textit{Top Left}: Semi-major axis. \textit{Top Right}:
  Orbital Eccentricity. \textit{Middle Left}: Primary's rotation
  period. \textit{Middle Right}: Primary's obliquity. \textit{Bottom
  Left}: Primary's Tidal Heat. \textit{Bottom Right}: The total
  angular momentum of the system.  }

\end{figure}

In Fig.~\ref{fig:ctl}, we show the history of \pr\ including solely the 
effects of CTL tidal evolution, once again for several choices of $\psi_P$. 
In this case, because there is much more energy in the orbit than
in the stellar rotation, $e$ and $a$ do not evolve much (in the case of 
low $\psi_P$) even though the stellar spins are evolving significantly.
In all cases the angular momentum is conserved to within a factor of 5
back to $t=-1$~Myr. If the system is evolved back further
than this, the requirement that the forcing frequency ($n$ in this
case) be nearly constant fails. The high $\psi_P$ case in general
conserves angular momentum poorly, due to the neglect of higher order 
$\psi$ terms which are especially important for $\psi \approx 90^\circ$.
Again, because the \pr\ system is presumed to have a nominal age of
$\approx$1~Myr, the calculations beyond $t=-1$~Myr are not reliable and
are shown only for context. Moreoever, the CTL model predictions for
high $\psi_P$ should be regarded with caution.
Finally, note that the behaviors shown here are modified in our final
treatment which includes the effects of radial contraction.

\bibliographystyle{apj}
\bibliography{par1802bib,par1802tides}

\begin{thebibliography}{55}
\expandafter\ifx\csname natexlab\endcsname\relax\def\natexlab#1{#1}\fi

\bibitem[{{Baraffe} {et~al.}(1998){Baraffe}, {Chabrier}, {Allard}, \&
  {Hauschildt}}]{bar98}
{Baraffe}, I., {Chabrier}, G., {Allard}, F., \& {Hauschildt}, P.~H. 1998, \aap,
  337, 403

\bibitem[{{Barnes} {et~al.}(2011){Barnes}, {Greenberg}, {Quinn}, {McArthur}, \&
  {Benedict}}]{barnes2011}
{Barnes}, R., {Greenberg}, R., {Quinn}, T.~R., {McArthur}, B.~E., \&
  {Benedict}, G.~F. 2011, \apj, 726, 71

\bibitem[{{Cargile} {et~al.}(2008){Cargile}, {Stassun}, \&
  {Mathieu}}]{rvsm2654}
{Cargile}, P.~A., {Stassun}, K.~G., \& {Mathieu}, R.~D. 2008, \apj, 674, 329

\bibitem[{{Carpenter} {et~al.}(2001){Carpenter}, {Hillenbrand}, \&
  {Skrutskie}}]{carp}
{Carpenter}, J.~M., {Hillenbrand}, L.~A., \& {Skrutskie}, M.~F. 2001, \aj, 121,
  3160

\bibitem[{{Chabrier} {et~al.}(2007){Chabrier}, {Gallardo}, \&
  {Baraffe}}]{cha07}
{Chabrier}, G., {Gallardo}, J., \& {Baraffe}, I. 2007, \aap, 472, L17

\bibitem[{{Covino} {et~al.}(2004){Covino}, {Frasca}, {Alcal{\'a}}, {Paladino},
  \& {Sterzik}}]{cov04}
{Covino}, E., {Frasca}, A., {Alcal{\'a}}, J.~M., {Paladino}, R., \& {Sterzik},
  M.~F. 2004, \aap, 427, 637

\bibitem[{{Covino} {et~al.}(2000){Covino}, {Catalano}, {Frasca}, {Marilli},
  {Fern{\'a}ndez}, {Alcal{\'a}}, {Melo}, {Paladino}, {Sterzik}, \&
  {Stelzer}}]{cov00}
{Covino}, E., {et~al.} 2000, \aap, 361, L49

\bibitem[{{D'Antona} \& {Mazzitelli}(1997)}]{dant}
{D'Antona}, F., \& {Mazzitelli}, I. 1997, Memorie della Societa Astronomica
  Italiana, 68, 807

\bibitem[{{Darwin}(1879)}]{Darwin1879}
{Darwin}, G.~H. 1879, Philosophical Transactions of the Royal Society, 170,
  447, (repr. Scientific Papers, Cambridge, Vol. II, 1908)

\bibitem[{{Darwin}(1880)}]{Darwin1880}
---. 1880, Royal Society of London Philosophical Transactions Series I, 171,
  713

\bibitem[{{Duval} {et~al.}(2004){Duval}, {Irace}, {Mainzer}, \&
  {Wright}}]{wise}
{Duval}, V.~G., {Irace}, W.~R., {Mainzer}, A.~K., \& {Wright}, E.~L. 2004, in
  Society of Photo-Optical Instrumentation Engineers (SPIE) Conference Series,
  Vol. 5487, Society of Photo-Optical Instrumentation Engineers (SPIE)
  Conference Series, ed. {J.~C.~Mather}, 101--111

\bibitem[{{Ferraz-Mello} {et~al.}(2008){Ferraz-Mello}, {Rodr{\'{\i}}guez}, \&
  {Hussmann}}]{FerrazMello2008}
{Ferraz-Mello}, S., {Rodr{\'{\i}}guez}, A., \& {Hussmann}, H. 2008, Celestial
  Mechanics and Dynamical Astronomy, 101, 171

\bibitem[{{Goldreich}(1966)}]{gold66}
{Goldreich}, P. 1966, \aj, 71, 1

\bibitem[{{Goldreich} \& {Soter}(1966)}]{GS1966}
{Goldreich}, P., \& {Soter}, S. 1966, Icarus, 5, 375

\bibitem[{{G{\'o}mez Maqueo Chew} {et~al.}(2009){G{\'o}mez Maqueo Chew},
  {Stassun}, {Pr{\v s}a}, \& {Mathieu}}]{sm4147}
{G{\'o}mez Maqueo Chew}, Y., {Stassun}, K.~G., {Pr{\v s}a}, A., \& {Mathieu},
  R.~D. 2009, \apj, 699, 1196

\bibitem[{{Greenberg}(2009)}]{Greenberg2009}
{Greenberg}, R. 2009, \apjl, 698, L42

\bibitem[{{Hauschildt} {et~al.}(1999){Hauschildt}, {Allard}, \&
  {Baron}}]{haus99}
{Hauschildt}, P.~H., {Allard}, F., \& {Baron}, E. 1999, \apj, 512, 377

\bibitem[{{Hebb} {et~al.}(2010){Hebb}, {Stempels}, {Aigrain},
  {Collier-Cameron}, {Hodgkin}, {Irwin}, {Maxted}, {Pollacco}, {Street},
  {Wilson}, \& {Stassun}}]{hebb10}
{Hebb}, L., {et~al.} 2010, \aap, 522, A37+

\bibitem[{{Heller} {et~al.}(2010){Heller}, {Jackson}, {Barnes}, {Greenberg}, \&
  {Homeier}}]{Heller2010}
{Heller}, R., {Jackson}, B., {Barnes}, R., {Greenberg}, R., \& {Homeier}, D.
  2010, \aap, 514, A22+

\bibitem[{{Heller} {et~al.}(2011){Heller}, {Leconte}, \& {Barnes}}]{Heller2011}
{Heller}, R., {Leconte}, J., \& {Barnes}, R. 2011, \aap, 528, A27+

\bibitem[{{Hillenbrand}(1997)}]{hill97}
{Hillenbrand}, L.~A. 1997, \aj, 113, 1733

\bibitem[{{Honeycutt}(1992)}]{honey92}
{Honeycutt}, R.~K. 1992, \pasp, 104, 435

\bibitem[{{Horne} \& {Baliunas}(1986)}]{horne}
{Horne}, J.~H., \& {Baliunas}, S.~L. 1986, \apj, 302, 757

\bibitem[{{Hut}(1981)}]{Hut1981}
{Hut}, P. 1981, \aap, 99, 126

\bibitem[{{Irwin} {et~al.}(2007){Irwin}, {Aigrain}, {Hodgkin}, {Stassun},
  {Hebb}, {Irwin}, {Moraux}, {Bouvier}, {Alapini}, {Alexander}, {Bramich},
  {Holtzman}, {Mart{\'{\i}}n}, {McCaughrean}, {Pont}, {Verrier}, \& {Zapatero
  Osorio}}]{jw380}
{Irwin}, J., {et~al.} 2007, \mnras, 380, 541

\bibitem[{{Jackson} {et~al.}(2009){Jackson}, {Barnes}, \& {Greenberg}}]{jack09}
{Jackson}, B., {Barnes}, R., \& {Greenberg}, R. 2009, \apj, 698, 1357

\bibitem[{{Kallrath} \& {Milone}(2009)}]{kall09}
{Kallrath}, J., \& {Milone}, E.~F. 2009, {Eclipsing Binary Stars: Modeling and
  Analysis}, ed. E.~F. Kallrath, J.~\&~Milone

\bibitem[{{Khaliullin} \& {Khaliullina}(2011)}]{KhaliullinKhaliullina2011}
{Khaliullin}, K.~F., \& {Khaliullina}, A.~I. 2011, \mnras, 411, 2804

\bibitem[{{Kirkpatrick} {et~al.}(1991){Kirkpatrick}, {Henry}, \&
  {McCarthy}}]{kirkpatrick91}
{Kirkpatrick}, J.~D., {Henry}, T.~J., \& {McCarthy}, Jr., D.~W. 1991, \apjs,
  77, 417

\bibitem[{{Leconte} {et~al.}(2010){Leconte}, {Chabrier}, {Baraffe}, \&
  {Levrard}}]{Leconte2010}
{Leconte}, J., {Chabrier}, G., {Baraffe}, I., \& {Levrard}, B. 2010, \aap, 516,
  A64+

\bibitem[{{Lin} {et~al.}(1996){Lin}, {Bodenheimer}, \& {Richardson}}]{lin96}
{Lin}, D.~N.~C., {Bodenheimer}, P., \& {Richardson}, D.~C. 1996, \nat, 380, 606

\bibitem[{{L{\'o}pez-Morales}(2007)}]{lop07}
{L{\'o}pez-Morales}, M. 2007, \apj, 660, 732

\bibitem[{{Luhman}(1999)}]{luh99}
{Luhman}, K.~L. 1999, \apj, 525, 466

\bibitem[{{Mathieu} {et~al.}(2007){Mathieu}, {Baraffe}, {Simon}, {Stassun}, \&
  {White}}]{mat07}
{Mathieu}, R.~D., {Baraffe}, I., {Simon}, M., {Stassun}, K.~G., \& {White}, R.
  2007, Protostars and Planets V, 411

\bibitem[{{Mazeh}(2008)}]{Mazeh2008}
{Mazeh}, T. 2008, in EAS Publications Series, Vol.~29, EAS Publications Series,
  ed. {M.-J.~Goupil \& J.-P.~Zahn}, 1--65

\bibitem[{{Murray} \& {Dermott}(1999)}]{murr99}
{Murray}, C.~D., \& {Dermott}, S.~F. 1999, {Solar system dynamics}, ed.
  {Murray, C.~D.~\& Dermott, S.~F.}

\bibitem[{{O'Dell} \& {Henney}(2008)}]{odell08}
{O'Dell}, C.~R., \& {Henney}, W.~J. 2008, \aj, 136, 1566

\bibitem[{{Palla} \& {Stahler}(1999)}]{PS99}
{Palla}, F., \& {Stahler}, S.~W. 1999, \apj, 525, 772

\bibitem[{{Pr{\v s}a} \& {Zwitter}(2005)}]{phoebe}
{Pr{\v s}a}, A., \& {Zwitter}, T. 2005, \apj, 628, 426

\bibitem[{{Scargle}(1982)}]{scargle}
{Scargle}, J.~D. 1982, \apj, 263, 835

\bibitem[{{Schwarzenberg-Czerny}(1991)}]{sigmap}
{Schwarzenberg-Czerny}, A. 1991, \mnras, 253, 198

\bibitem[{{Siess} {et~al.}(2000){Siess}, {Dufour}, \& {Forestini}}]{SDF00}
{Siess}, L., {Dufour}, E., \& {Forestini}, M. 2000, \aap, 358, 593

\bibitem[{{Simon} \& {Obbie}(2009)}]{simon09}
{Simon}, M., \& {Obbie}, R.~C. 2009, \aj, 137, 3442

\bibitem[{{Stassun} {et~al.}(2008){Stassun}, {Mathieu}, {Cargile}, {Aarnio},
  {Stempels}, \& {Geller}}]{natsm2654}
{Stassun}, K.~G., {Mathieu}, R.~D., {Cargile}, P.~A., {Aarnio}, A.~N.,
  {Stempels}, E., \& {Geller}, A. 2008, \nat, 453, 1079

\bibitem[{{Stassun} {et~al.}(1999){Stassun}, {Mathieu}, {Mazeh}, \&
  {Vrba}}]{sta99}
{Stassun}, K.~G., {Mathieu}, R.~D., {Mazeh}, T., \& {Vrba}, F.~J. 1999, \aj,
  117, 2941

\bibitem[{{Stassun} {et~al.}(2006){Stassun}, {Mathieu}, \& {Valenti}}]{nt06}
{Stassun}, K.~G., {Mathieu}, R.~D., \& {Valenti}, J.~A. 2006, \nat, 440, 311

\bibitem[{{Stassun} {et~al.}(2007){Stassun}, {Mathieu}, \& {Valenti}}]{pii}
---. 2007, \apj, 664, 1154

\bibitem[{{Stassun} {et~al.}(2004){Stassun}, {Mathieu}, {Vaz}, {Stroud}, \&
  {Vrba}}]{sta04}
{Stassun}, K.~G., {Mathieu}, R.~D., {Vaz}, L.~P.~R., {Stroud}, N., \& {Vrba},
  F.~J. 2004, \apjs, 151, 357

\bibitem[{{Stassun} {et~al.}(2002){Stassun}, {van den Berg}, {Mathieu}, \&
  {Verbunt}}]{sta02}
{Stassun}, K.~G., {van den Berg}, M., {Mathieu}, R.~D., \& {Verbunt}, F. 2002,
  \aap, 382, 899

\bibitem[{{Stempels} {et~al.}(2008){Stempels}, {Hebb}, {Stassun}, {Holtzman},
  {Dunstone}, {Glowienka}, \& {Frandsen}}]{asas08}
{Stempels}, H.~C., {Hebb}, L., {Stassun}, K.~G., {Holtzman}, J., {Dunstone},
  N., {Glowienka}, L., \& {Frandsen}, S. 2008, \aap, 481, 747

\bibitem[{{Stempels} \& {Piskunov}(2003)}]{stem03}
{Stempels}, H.~C., \& {Piskunov}, N. 2003, \aap, 408, 693

\bibitem[{{Torres} \& {Ribas}(2002)}]{torr02}
{Torres}, G., \& {Ribas}, I. 2002, \apj, 567, 1140

\bibitem[{{Whitney} {et~al.}(2003){Whitney}, {Wood}, {Bjorkman}, \&
  {Cohen}}]{whit03}
{Whitney}, B.~A., {Wood}, K., {Bjorkman}, J.~E., \& {Cohen}, M. 2003, \apj,
  598, 1079

\bibitem[{{Zahn}(2008)}]{Zahn2008}
{Zahn}, J.-P. 2008, in EAS Publications Series, Vol.~29, EAS Publications
  Series, ed. {M.-J.~Goupil \& J.-P.~Zahn}, 67--90

\bibitem[{{Zahn} \& {Bouchet}(1989)}]{zahn1989}
{Zahn}, J.-P., \& {Bouchet}, L. 1989, \aap, 223, 112

\end{thebibliography}

\clearpage

\begin{deluxetable}{lccr}
\tabletypesize{\small}
\tablecolumns{4}
\tablewidth{0pc}
\tablecaption{\textsc{Photometric Time Series Observations of \pr}}
\tablehead{
	\colhead{Telescope} & 
	\colhead{HJD Range\tablenotemark{a}} &
	\colhead{Filter} & 
	\colhead{N$_{obs}$\tablenotemark{b} } 
} 

\startdata
KPNO 0.9-m &    49698.35--49714.50 &    \ic &   110 \\ 
KPNO 0.9-m &    50820.62--50829.78 &    \ic &   21  \\ 
CTIO 0.9-m &    51929.59--51936.78 &    \ic &   164 \\ 
\  &          \ &                   $V$\ &   153 \\
KPNO 0.9-m &    52227.75--52238.00&     \ic &   131 \\ 
KPNO 0.9-m &    52595.75--52624.95&     \ic &   279 \\ 
\  &          \ &                   $V$\ &   146 \\
CTIO 0.9-m &    52622.57--52631.51&     \ic &   80      \\
\  &          \ &                   $V$\ &    83 \\
SMARTS 0.9-m &    53011.57--53024.77&     \ic &   200 \\ 
\  &          \ &                   $V$\ &   104 \\ 
SMARTS 1.3-m &  53403.53--53463.53&     \ic &   246 \\ 
\  &          \  &                  $V$\ &   176 \\
\  &          \  &                  $J$\ &    90 \\
\  &          \  &                  $K_C$\ &  88 \\
SMARTS 1.3-m &  53646.86--53728.69&     \ic &   188 \\ 
\  &          \  &                  $V$\ &   113 \\
\  &          \  &                  $J$\ &    57 \\
\  &          \  &                  $K_C$\ &  52 \\
SMARTS 1.0-m & 53719.56--53727.83& \ic & 117 \\ %
\  &          \  &                  $V$\ &  101 \\ 
SMARTS 1.3-m &  53745.63--53846.51&     \ic &   276 \\ 
\  &          \  &                  $V$\ &   182 \\
\  &          \  &                  $J$\ &    80 \\
\  &          \  &                  $K_C$\ &  73 \\
SMARTS 1.3-m &  53980.89--54100.65 &    \ic &   254 \\ 
\  &          \  &                  $V$\ &   190 \\
\  &          \  &                  $J$\ &    99 \\
\  &          \  &                  $K_C$\ &  98 \\
SMARTS 1.0-m & 54103.58--54112.773& \ic & 105 \\ %
\  &          \  &                  $V$\ &  103 \\ 
SMARTS 1.3-m &  54103.73--54191.53&     \ic &   183 \\ 
\  &          \  &                  $V$\ &   61 \\
\  &          \  &                  $J$\ &    63 \\
\  &          \  &                  $K_C$\ &  54 \\
SMARTS 1.3-m &  54375.81--54465.82&     \ic &   371 \\ 
\  &          \  &                  $V$\ &   250 \\
\  &          \  &                  $J$\ &   128 \\
\  &          \  &                  $H$\ &   129 \\
SMARTS 1.3-m &  54467.62--54497.69&     \ic &   142 \\ 
\  &          \  &                  $V$\ &   96 \\
\  &          \  &                  $J$\ &    47 \\
\  &          \  &                  $H$\ &    47 \\
SMARTS 1.0-m &  54482.58--54494.74&    \ic & 218 \\ 
\  &          \  &                  $V$\ &    169 \\
\  &          \  &                  $B$\ &   183 \\ 
SMARTS 1.0-m &  54835.56--54853.78&     \ic &   403 \\ 
\  &          \  &                  $V$\ &   359 \\

\enddata

\label{table:obs}
\tablenotetext{a}{Range of Heliocentric Julian Dates (2\,400\,000+).}
\tablenotetext{b}{Number of observations.}
\end{deluxetable}

\begin{deluxetable}{lrr}
\tablecaption{\textsc{Differential $V$-band Light Curve of \pr}}
\tablewidth{0pc}
\tablehead{\colhead{HJD\tablenotemark{a}} &  \colhead{$\Delta$$m$\tablenotemark{b}} & \colhead{$\sigma_m$}}
\startdata
51930.557737  &     0.006  &     0.010  \\
51930.569416  &    -0.001  &     0.010  \\
51930.581025  &     0.007  &     0.010  \\
51930.592365  &    -0.007  &     0.010  \\
51930.603794  &     0.007  &     0.010  \\
51930.615404  &     0.000  &     0.010  \\
51930.626933  &     0.001  &     0.010  \\
51930.638092  &    -0.005  &     0.010  \\
51930.656051  &    -0.001  &     0.010  \\
51930.667491  &     0.001  &     0.011  \\
51930.679160  &     0.006  &     0.011  \\
51930.693249  &    -0.006  &     0.010  \\
51930.704649  &     0.007  &     0.010  \\
51930.735757  &     0.003  &     0.011  \\
51930.748306  &    -0.008  &     0.011  \\
\enddata
\tablenotetext{a}{\footnotesize Heliocentric Julian Date (2\,400\,000+).}
\tablenotetext{b}{\footnotesize Differential $V$ magnitude}
\label{table:lcv}
\tablecomments{\footnotesize This table is published in its entirety in a machine-readable form in the online journal.  A portion is shown here for guidance regarding its form and content.}
\end{deluxetable}

\begin{deluxetable}{lrr}
\tablecaption{\textsc{Differential \ic-band Light Curve of \pr}}
\tablewidth{0pc}
\tablehead{\colhead{HJD\tablenotemark{a}} &  \colhead{$\Delta$$m$\tablenotemark{b}} & \colhead{$\sigma_m$}}
\startdata
49701.860452  &    -0.040  &     0.020  \\
49701.913182  &    -0.026  &     0.020  \\
49701.938571  &    -0.026  &     0.020  \\
49701.976661  &     0.006  &     0.020  \\
49702.006931  &    -0.001  &     0.020  \\
49702.687600  &    -0.012  &     0.020  \\
49702.716890  &    -0.015  &     0.020  \\
49702.745210  &     0.003  &     0.020  \\
49702.773530  &     0.007  &     0.020  \\
49702.805760  &     0.009  &     0.020  \\
49702.861430  &     0.028  &     0.020  \\
49702.889750  &     0.027  &     0.020  \\
49702.918070  &     0.031  &     0.020  \\
49702.950290  &     0.032  &     0.020  \\
49702.981540  &     0.030  &     0.020  \\
\enddata
\tablenotetext{a}{\footnotesize Heliocentric Julian Date (2\,400\,000+).}
\tablenotetext{b}{\footnotesize Differential \ic\ magnitude}
\tablecomments{\footnotesize This table is published in its entirety in a machine-readable form in the online journal.  A portion is shown here for guidance regarding its form and content.}
\label{table:lci}
\end{deluxetable}

\begin{deluxetable}{lr}
\tablecaption{\textsc{Differential $J$-band Light Curve of \pr}}
\tablewidth{0pc}
\tablehead{\colhead{HJD\tablenotemark{a}} &  \colhead{$\Delta$$m$\tablenotemark{b}}}
\startdata
54013.794832  &    -0.010  \\
54040.717805  &    -0.002  \\
54041.709868  &     0.012  \\
54005.796854  &    -0.006  \\
53981.864644  &     0.018  \\
53999.786555  &     0.002  \\
54071.679609  &     0.150  \\
54002.846548  &    -0.025  \\
54003.778756  &     0.036  \\
54019.797322  &    -0.008  \\
54020.805072  &     0.013  \\
53993.868102  &    -0.027  \\
54023.779094  &     0.025  \\
54049.717259  &     0.008  \\
54050.713167  &     0.136  \\
54024.794247  &    -0.021  \\
\enddata
\tablenotetext{a}{\footnotesize Heliocentric Julian Date (2\,400\,000+).}
\tablenotetext{b}{\footnotesize Differential $J$ magnitude}
\tablecomments{\footnotesize This table is published in its entirety in a machine-readable form in the online journal.  A portion is shown here for guidance regarding its form and content.}
\label{table:lcj}
\end{deluxetable}

\begin{deluxetable}{lr}
\tablecaption{\textsc{Differential $H$-band Light Curve of \pr}}
\tablewidth{0pc}
\tablehead{\colhead{HJD\tablenotemark{a}} &  \colhead{$\Delta$$m$\tablenotemark{b}}}
\startdata
54376.787242  &    -0.014  \\
54377.789689  &     0.143  \\
54378.775544  &    -0.020  \\
54378.785284  &     0.012  \\
54380.739724  &    -0.019  \\
54381.776602  &    -0.025  \\
54381.886711  &    -0.012  \\
54382.748298  &     0.008  \\
54383.729640  &     0.004  \\
54383.859218  &     0.005  \\
54384.742592  &     0.047  \\
54384.862200  &     0.122  \\
54385.741196  &    -0.008  \\
54385.851496  &    -0.004  \\
\enddata
\tablenotetext{a}{\footnotesize Heliocentric Julian Date (2\,400\,000+).}
\tablenotetext{b}{\footnotesize Differential $H$ magnitude}
\tablecomments{\footnotesize This table is published in its entirety in a machine-readable form in the online journal.  A portion is shown here for guidance regarding its form and content.}
\label{table:lch}
\end{deluxetable}

\begin{deluxetable}{lr}
\tablecaption{\textsc{Differential $K_S$-band Light Curve of \pr}}
\tablewidth{0pc}
\tablehead{\colhead{HJD\tablenotemark{a}} &  \colhead{$\Delta$$m$\tablenotemark{b}}}
\startdata
54013.798334  &    -0.006  \\
54040.721410  &     0.005  \\
54041.713265  &     0.001  \\
54005.800356  &    -0.011  \\
53981.868447  &     0.015  \\
53999.790149  &     0.012  \\
54071.683111  &     0.151  \\
54002.850061  &    -0.017  \\
54003.782373  &     0.043  \\
54019.800928  &    -0.002  \\
54020.808574  &    -0.009  \\
53993.871639  &    -0.011  \\
54023.782514  &     0.027  \\
54049.720749  &     0.003  \\
54050.716645  &     0.133  \\
\enddata
\tablenotetext{a}{\footnotesize Heliocentric Julian Date (2\,400\,000+).}
\tablenotetext{b}{\footnotesize Differential $K_S$ magnitude}
\tablecomments{\footnotesize This table is published in its entirety in a machine-readable form in the online journal.  A portion is shown here for guidance regarding its form and content.}
\label{table:lck}
\end{deluxetable}

\begin{deluxetable}{cccl}
\tabletypesize{\small}
\tablecolumns{5}
\tablewidth{0pc}
\tablecaption{\textsc{Timings of Eclipse Minima in the \ic\ Light Curve}}

\tablehead{
        \colhead{\textsc{HJD}\tablenotemark{a}} &
        \colhead{\textsc{O--C (Phase)}} &
        \colhead{\textsc{Eclipse Type}} & \colhead{\textsc{Telescope}}        }

\startdata
    49701.567326        $\pm$ 0.000006 &   -0.007339 $\pm$       0.000001&  Secondary &  KPNO 0.9-m \\
    49703.956710        $\pm$ 0.000005 &    0.004684 $\pm$       0.000001&Primary &  KPNO 0.9-m \\
    49713.296386        $\pm$ 0.000001 &    0.0029800 $\pm$      0.0000002& Primary &  KPNO 0.9-m \\
    51935.7554          $\pm$ 0.0001 &    0.00590       $\pm$   0.00002& Secondary &  CTIO 0.9-m \\
    52227.86081         $\pm$ 0.00004   &   0.003555 $\pm$       0.000009& Primary &  KPNO 0.9-m \\
    52234.84326         $\pm$ 0.00001   &  -0.002650 $\pm$        0.000002& Secondary  &  KPNO 0.9-m  \\
    52601.77110         $\pm$ 0.00008 &     0.00311     $\pm$   0.00002& Primary &  KPNO 0.9-m \\
    52622.76956         $\pm$ 0.00002 &   -0.004397 $\pm$  0.000004&Secondary     &  KPNO 0.9-m, CTIO 0.9-m  \\
    52629.82635         $\pm$ 0.00002 &    0.006356 $\pm$       0.000004&  Primary &  CTIO 0.9-m \\
    53017.71700         $\pm$ 0.00005 &   -0.00291     $\pm$   0.00001&  Primary &  SMARTS 0.9-m \\
    53024.74611         $\pm$ 0.00001 &    0.000075 $\pm$       0.000002& Secondary    &  SMARTS 0.9-m    \\ 
    54106.74615         $\pm$ 0.00001 &   -0.001214 $\pm$       0.000002&Primary &  SMARTS 1.0-m, 1.3-m  \\
    54487.65947         $\pm$ 0.00007 &   -0.00369     $\pm$   0.00002& Secondary  &   SMARTS 1.0-m, 1.3-m    \\ 
    54494.6722          $\pm$ 0.0005    &   -0.0030      $\pm$   0.0001&Primary &  SMARTS 1.0-m, 1.3-m \\
    54847.563143        $\pm$ 0.000004 &   -0.0011578 $\pm$      0.0000009& Secondary & SMARTS 1.0-m       \\
\enddata
\tablenotetext{a}{Heliocentric Julian Date (2\,400\,000+)}
\label{table:timings}
\end{deluxetable}

\begin{deluxetable}{lcccc}
\tabletypesize{\small}
\tablecolumns{5}
\tablewidth{0pc}
\tablecaption{\textsc{Periodicity in the Light Curves of \pr\ in Days}}

\tablehead{
	\phn  & 
	\multicolumn{2}{c}{\textsc{OFE}\tablenotemark{a}} & 
	\multicolumn{2}{c}{\textsc{O--C}\tablenotemark{b}} \\
	\colhead{\textsc{Passband}} & 
	\colhead{$P_1$} &
        \colhead{$P_2$} & 
        \colhead{$P_1$} &
        \colhead{$P_2$} }

\startdata

$V$\dotfill & 
	4.626 $\pm$ 0.001 & 0.73557 $\pm$  0.00002 &  4.6257 $\pm$ 0.0009& 0.73558 $\pm$ 0.00001 \\	
$I_C$\dotfill &
	4.6257 $\pm$ 0.0005& 0.73560 $\pm$ 0.00001& 4.6259 $\pm$ 0.0004& 0.735606 $\pm$ 0.000009\\
$J$\dotfill & 
	4.628 $\pm$ 0.003& 0.73551 $\pm$ 0.00007& 4.627 $\pm$ 0.002& 0.73551 $\pm$ 0.00005\\
$H$\dotfill & 
	4.64 $\pm$ 0.03& 0.7353 $\pm$ 0.0008& 4.64 $\pm$ 0.03& 0.7353 $\pm$ 0.0007\\
$K_S$\dotfill & 
	4.629 $\pm$ 0.003& 0.7355 $\pm$ 0.0001& 4.627 $\pm$ 0.004& 0.7355 $\pm$ 0.0001\\

\enddata

\tablenotetext{a}{Only the phases of the light curves that are out-of-eclipse, i.e., excluding the eclipses, 
were searched for periodicities.}
\tablenotetext{b}{We did the periodicity analysis on the residuals of the modeling of the light curves; 
any periodicity due to the EB nature of the system would be removed from the O--C Periodograms.}
\label{table:periods}
\end{deluxetable}

\begin{deluxetable}{lcc}
\tabletypesize{\small}
\tablecolumns{3}
\tablewidth{0pc}
\tablecaption{\textsc{Amplitude of Periodic Photometric Variability of \pr}}
\tablehead{\textsc{Passband} & \textsc{$A_{P_1}$} & \textsc{$A_{P_2}$}\\
\phn & (mag) & (mag)
}

\startdata

$V$  \dotfill 
	&   0.029 
	&   0.016 \\  	
$I_C$ \dotfill
	&          0.016 
	&     0.015  \\
$J$   \dotfill
	&       0.011 
	&   0.009     \\ 
$H$  \dotfill
	&       0.012 
	&   0.013 \\
$K_S$ \dotfill  &     0.009 
	&     0.011  \\

\enddata
\label{table:amps}
\end{deluxetable}

\begin{deluxetable}{lccc}
\tabletypesize{\small}
\tablecolumns{4}
\tablewidth{0pc}
\tablecaption{\textsc{Orbital and Physical Parameters of \pr\label{table:params}}}
\tablehead{
\colhead{\phn}&\colhead{\phn}&
\colhead{ \pii} &
\colhead{ This work} \\ 
\colhead{\phn}&\colhead{\phn}&
\colhead{\small RVs $+$ \ic} & 
\colhead{\small RVs $+$ \op\nir}  
}

\startdata
Orbital period  &  $P_{\rm orb}$ (d)        &   4.673843 $\pm$ 0.000068&        4.673903 $\pm$ 0.000060 \\
Epoch of primary minimum\tablenotemark{a} & HJD$_0$ (d) &  \nodata    &  54849.9008 $\pm$ 0.0005   \\
Eccentricity  &  {\it e}                        &0.029 $\pm$ 0.005 &            0.0166 $\pm$ 0.003\\
Orientation  of periastron  &  $\omega$ ($\pi$ rads)    &1.478 $\pm$ 0.010 &    1.484 $\pm$ 0.010\\
Semi-major axis  &  $a\sin{i}$ (AU)
                        &0.0501 $\pm$ 0.0006
                        &0.0496 $\pm$ 0.0008\tablenotemark{\dagger}\\
Inclination angle  &  $i$ (\degr)
                        &78.1 $\pm$ 0.6
                        &80.8 $^{+\ 8.0}_{-\ 2.0}$\tablenotemark{\ddag}\\
Systemic velocity  &  $v_\gamma$ (km s$^{-1}$)
                        &23.7 $\pm$ 0.5
                        &23.4 $\pm$ 0.7\tablenotemark{\dagger}\\
Primary semiamplitude  &  $K_1$ (km s$^{-1}$)
                        &57.74 $\pm$ 0.75\tablenotemark{b}
                        &57.28 $\pm$ 2.20 \\ 
Secondary semiamplitude  &  $K_2$ (km s$^{-1}$)
                        &58.92 $\pm$ 0.95\tablenotemark{b}
                        &58.19 $\pm$ 2.78 \\ 
Mass ratio  &  $q \equiv M_2/M_1$
                        &0.98 $\pm$ 0.01
                        &0.985 $\pm$ 0.029\tablenotemark{\dagger}\\
Total mass  &  $M\sin^3{i}$ (M$_\sun$)
                        &0.768 $\pm$ 0.028
                        &0.745 $\pm$ 0.034\tablenotemark{\dagger}\\
Primary mass  &  $M_1$ (M$_\sun$)
                        &0.414 $\pm$ 0.015
                        &0.391 $\pm$ 0.032\\
Secondary mass  &  $M_2$ (M$_\sun$)
                        &0.406 $\pm$ 0.014
                        &0.385 $\pm$ 0.032\\
Primary radius  &  $R_1$ (R$_\sun$)
                        &1.82 $\pm$ 0.05
                        &1.73 $^{+\ 0.01}_{-\ 0.02}$\tablenotemark{\ddag}\\
Secondary radius  &  $R_2$ (R$_\sun$)
                        &1.69 $\pm$ 0.018
                        &1.62 $^{+\ 0.01}_{-\ 0.02}$\tablenotemark{\ddag}\\
Primary gravity  &  $\log{g_1}$
                        &3.54 $\pm$ 0.09\tablenotemark{b}
                        &3.55 $\pm$ 0.04\\
Secondary gravity  &  log $g_2$
                        &3.62 $\pm$ 0.10\tablenotemark{b}
                        &3.61 $\pm$ 0.04\\
Primary surface potential  &  $\Omega_1$
                        &\nodata
                        &7.27 $\pm$ 0.06 \\
Secondary surface potential  &  $\Omega_2$
                        &\nodata
                        &7.62 $\pm$ 0.06 \\
Primary synchronicity parameter  &  $F_{1}$ (d$^{-1}$)
                        &\nodata 
                        &1.0097 $\pm$ 0.0013 \\
Secondary synchronicity parameter  &  $F_{2}$ (d$^{-1}$)
                        &\nodata        
                        &1.0097 $\pm$ 0.0013 \\
Effective temperature ratio  &  $T_{\rm eff,1}/T_{\rm eff,2}$
                        &1.084 $\pm$ 0.007
                        &1.0924 $\pm$ 0.0017\\
Primary effective temperature & $T_{\rm eff,1}$ (K) & 3945 $\pm$ 100 & 3675 $\pm$ 150\tablenotemark{\ast}\\
Secondary effective temperature & $T_{\rm eff,2}$ (K) & 3655 $\pm$ 100 & 3365 $\pm$ 150\tablenotemark{\ast}\\

\enddata

\tablenotetext{a}{\footnotesize Heliocentric Julian Date (2\,400\,000+).}
\tablenotetext{b}{\footnotesize Calculated from parameters and uncertainties in \pii.}
\tablenotetext{\dagger}{\footnotesize
The uncertainties in these parameters are conservatively estimated from the formal errors of a fit to the RV data alone. See \S\ref{eb}.}
\tablenotetext{\ddag}{\footnotesize
The uncertainties in these parameters are conservatively estimated from a variation in the level of third light between 5 and 75\% of the system's total luminosity. See \S\ref{eb}.}
\tablenotetext{\ast}{\footnotesize The uncertainty in \teff\ is
dominated by the systematic uncertainty in the conversion to \teff\
from the mean spectral type that we adopt for the system (see \S\ref{eb}).
The \teff\ ratio, via which $T_{\rm eff,2}$ is derived from \teff $_{\rm ,1}$, is
independently and accurately determined from the light curves;
$T_{\rm eff,1}$ and $T_{\rm eff,2}$ differ by 9.2$\pm$0.2\% regardless
of their absolute value.}
\end{deluxetable}

\begin{deluxetable}{ccccc}
\tablecolumns{5}
\tablecaption{\textsc{Fits Parameters to Radial Contraction Models\label{tab:rory}}}
\tablewidth{0pc}
\tablehead{
\colhead{Model} & \colhead{$a_0$} & \colhead{$a_1$} & \colhead{$a_2$} & \colhead{$a_3$}
}
\startdata
\citet{bar98} & $-$1.754 & 1.378 & 0.3444 & 0.02758\\
\citet{dant} & $-$2.557 & 2.7085 & 0.9177 & 0.09971\\
\enddata
\end{deluxetable}

\end{document}